\documentclass[aps,prl,twocolumn,showpacs,superscriptaddress,reprint]{revtex4-1}
\usepackage{amsmath}
\usepackage{amsfonts}
\usepackage{amssymb}
\usepackage{bm}
\usepackage[usenames,dvipsnames]{color} 
\usepackage[dvipdfmx]{graphicx}
\usepackage{epstopdf}
\usepackage{ulem}

\usepackage{graphicx}

\begin{document}

\title{Ideal glass states are not purely vibrational: Insight from randomly pinned glasses}

\author{Misaki Ozawa}
\affiliation{Laboratoire Charles Coulomb (L2C), 
University of Montpellier and CNRS, F-34095 Montpellier, France}

\author{Atsushi Ikeda}
\affiliation{Graduate School of Arts and Sciences, University of Tokyo, 
Tokyo, 3-8-1, Japan}

\author{Kunimasa Miyazaki}
\affiliation{Department of Physics, Nagoya
University, Nagoya 464-8602, Japan}

\author{Walter Kob}
\email[Corresponding author: ]{walter.kob@umontpellier.fr}\affiliation{Laboratoire Charles Coulomb (L2C),
University of Montpellier and CNRS, F-34095 Montpellier, France}

\date{\today}

\begin{abstract}

We use computer simulations to probe the thermodynamic and dynamic
properties of a glass-former that undergoes an ideal glass-transition
because of the presence of randomly pinned particles. We find that
even deep in the equilibrium glass state the system relaxes to some
extent because of the presence of localized excitations that allow
the system to access different inherent structures, giving thus rise
to a non-trivial contribution to the entropy. By calculating with high
accuracy the vibrational part of the entropy, we show that also in the
equilibrium glass state thermodynamics and dynamics give a coherent
picture and that glasses should not be seen as a disordered solid in
which the particles undergo just vibrational motion but instead as a 
system with a highly nonlinear internal dynamics.

\end{abstract}

\pacs{Valid PACS appear here}

\maketitle

The nature of the glass transition is one of
the most challenging research topics in condensed
matter physics and has therefore been in the focus of a multitude of
studies~\cite{binder2011glassy,berthier2011theoretical,cavagna2009supercooled,dyre2006colloquium,chandler2009dynamics}.
Many aspects of the slow and complex dynamics of
supercooled liquids and the resulting glass transition have
been successfully explained in terms of the potential energy landscape
(PEL)~\cite{goldstein1969viscous,stillinger1982hidden,wales2003energy,sciortino2005potential,heuer2008exploring}.
In this framework the configurational space of the
system is partitioned into basins of
attractions of the local energy minima (the inherent structures, IS) of
the potential energy and the dynamics at low temperatures is characterized
as the motion through the complex pathway that connects neighboring
basins. The conventional description of glasses is that the glass
state corresponds to a vibrational motion around the IS, which in real
space means that the particles are trapped by the cages formed by their
neighbors and vibrate around a fixed amorphous configuration. However,
recent experiments challenge this view since they seem to suggest
that certain types of relaxation processes are present even in the
glass, implying that the motion of the atoms is more complex than pure
vibrations~\cite{ruta2012atomic,bock2013cooperative,giordano2016unveiling}.
However, it is difficult to decide whether or not these relaxation
processes are indeed an equilibrium property of the sample or just
related to aging. The existence of equilibrium relaxation processes
in the glass state implies that even at low temperatures the system
explores a complex landscape, i.e., it can access in a {\it finite} time
many different local minima. This in turn has the consequence that the
conventional definition of the configurational entropy $s_{\rm conf}$
as the difference between the total entropy $s_{\rm tot}$ of the system
and its purely vibrational part $s_{\rm vib}$ should be questioned.  Thus the
study of (equilibrium!) relaxation processes in the equilibrium glass
state will allow to advance our understanding of the meaning of the
glass state on the microscopic level.

Advancing on this question has so far been hampered by the fact
that it was impossible to generate equilibrium glasses (also called
``ideal glasses''), i.e., {\it equilibrium} structures at very low
temperatures. This situation has recently changed since it has been
realized that if one pins (immobilizes) randomly a finite fraction $c$ of
the particles~\cite{kim2003effects}, the fluid particles, i.e., non-pinned
ones, undergo an equilibrium glass transition if $c$ is increased beyond
a certain threshold~\cite{cammarota2012ideal,berthier2012static}. 
Numerical simulations of simple glass-formers have confirmed that this
pinning approach does indeed allow to observe an equilibrium glass
transition at which the entropy shows a marked bend and hence to access
the equilibrium glass state~\cite{kob2013probing,ozawa2015equilibrium}.
Thus these results have opened the door to study the properties of
equilibrium glasses and in the following we will demonstrate that
even in the ideal glass state relaxation processes are present,
implying that in the glass the entropy is not just given by the
vibrational contribution.

We simulate a binary mixture of $N$ Lennard-Jones (LJ) particles
in three dimensions~\cite{kob1995testing1} of which a fraction $c$
are permanently pinned. $N$ is 300 or 1200 and we use the standard LJ
units for the length, energy, and temperature, setting the Boltzmann
constant $k_{\rm B}=1$. First we have equilibrated the system without
pinned particles at a given temperature $T$, then the positions of $cN$
particles are frozen and we study the static and dynamical properties
of the remaining $(1-c)N$ mobile particles of the system. To study the
static properties, we use the parallel tempering (PT) molecular dynamics
method with $24$ replicas~\cite{hukushima1996exchange,yamamoto2000replica}
(see Ref.~\cite{kob2013probing,ozawa2015equilibrium} for details on the
pinning procedure and the PT).  For the dynamical properties we use the
standard Monte Carlo (MC) dynamics simulation~\cite{berthier2007monte},
where an elementary move is a random displacement of a randomly chosen
particle within a linear box size $\delta= 0.15$.  One MC step, which is
our unit of time, consists of $(1-c)N$ such attempts.  Despite the huge
acceleration of the sampling due to the PT, we found that runs up to
$2 \cdot 10^{10}$ steps were needed to get statistically significant
results. This, and the necessary averaging over the independent
realizations of the pinned configurations (typically $25$ for $N=300$
and $10$ for $N=1200$).

\begin{figure}[tb]
\includegraphics[width=0.85\columnwidth]{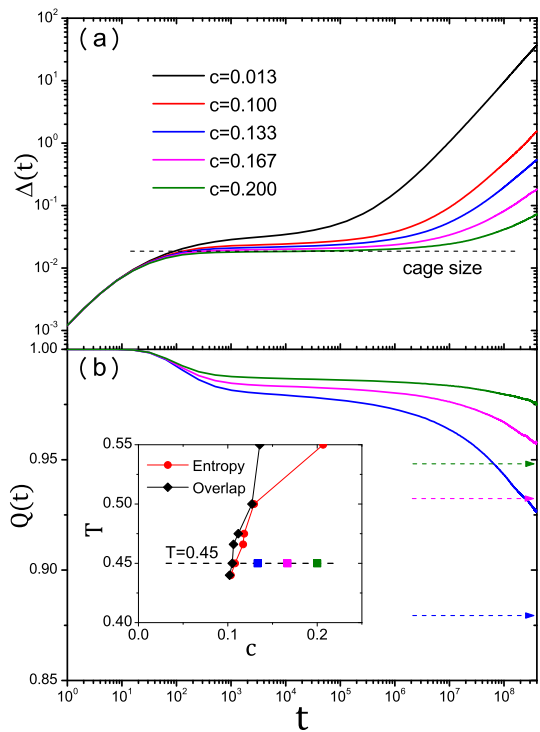}
\caption{
(a) Time dependence of the mean squared displacement $\Delta(t)$
for different concentrations of pinned particles, $c$, at $T=0.45$ for $N=1200$.
The horizontal dashed line is the height of the plateau for $c=0.2$.  (b)
The dynamic overlap function $Q(t)$ at $T=0.45$ for $N=1200$ and concentrations $c$
at which the system is in the equilibrium glass state.  The horizontal
dashed arrows are the static overlap functions $Q^{\rm (static)}$
evaluated from the parallel tempering simulations.  The inset shows the
state points for the presented $Q(t)$ in the phase diagram
from Ref.~\cite{ozawa2015equilibrium}.
}
\label{fig1:msd_overlap}
\end{figure}

To probe the relaxation dynamics of the system we focus on $T=0.45$
and increase $c$, thus crossing at around $c=0.1$ the boundary between
fluid and glass state (see inset of Fig.~\ref{fig1:msd_overlap}(b)).
Figure~\ref{fig1:msd_overlap}(a) shows the mean squared displacement
(MSD) for different values of $c$: $\Delta(t) = \sum_{i=1}^{(1-c)N}
[\langle |{\bf r}_i(t) - {\bf r}_i(0)|^2\rangle]/(1-c)N$. Here
$\langle \cdots \rangle$ and $[\cdots]$ are the thermal and disorder
averages, respectively. Note that we have averaged the MSD over
both species of particles. At intermediate times $\Delta(t)$ has a
marked plateau that is related to the usual cage effect observed in
glassy systems~\cite{binder2011glassy} and the horizontal dashed line
indicates the plateau height of $\Delta(t)$ at $c=0.2$. Surprisingly we
find that even in the equilibrium glass phase, i.e.,~at this temperature
for $c\geq 0.10$ (see SI), $\Delta(t)$ shows at long times a marked
increase above this plateau, indicating that the particles can leave
their cage.  To understand this behavior, we analyze the van-Hove
correlation function (see SI) and find that the particles undergo an
exchange motions, i.e.,  the particles tend to be replaced by the same
kind of particles (see SI).  This seems to explain the upturn of the
mean squared displacement.  More surprising is the results regarding the
collective overlap $Q(t) = \sum_{i,j} [\langle \theta (a - |{\bf r}_i(t)
- {\bf r}_j(0)|)\rangle]$, where $\theta(x)$ is the Heaviside function
and $a=0.3$.  This function probes the {\it collective} relaxation of
the system and it is not affected by the particle exchange motions.
Figure~\ref{fig1:msd_overlap}(b) shows that $Q(t)$ decays slightly
after the plateau at intermediate times, before it decays to its
long time limit (horizontal dashed lines $Q(t \to \infty)=Q^{\rm
(static)}$), a quantity that can be calculated with high accuracy
directly from the PT simulations.  The very high value of $Q^{\rm
(static)}>0.85$ and the strong $c-$dependence of this quantity, see
Ref.~\cite{ozawa2015equilibrium}, demonstrates that the explored state
points are indeed glass states and hence we can conclude that the system
shows subtle and non-trivial relaxation dynamics even in the ideal glass.
In the SI we show that this is {\it not} an out-of-equilibrium effect.

The presence of this relaxation dynamics is at odds with the results
of Ref.~\cite{ozawa2015equilibrium} that in the glass phase the
configurational entropy $s_{\rm conf}$ seems to be zero (see red circles
in Fig.~\ref{fig3:entropy}(b)), since $s_{\rm conf}=0$ implies that
there are no states into which the system can move to relax. In that
work $s_{\rm conf}$ was estimated from $s_{\rm tot}-s_{\rm harm}$,
where $s_{\rm harm}$ is the entropy of the strictly harmonic solid,
i.e., a quantity that can be obtained directly from the vibrational
density of states.  Our present results rise thus the question
whether $s_{\rm conf}$ can indeed be approximated reliably by this
difference~\cite{sciortino1999inherent,berthier2014novel,johari2002entropy}.

\begin{figure}[tb]
\includegraphics[width=0.80\columnwidth]{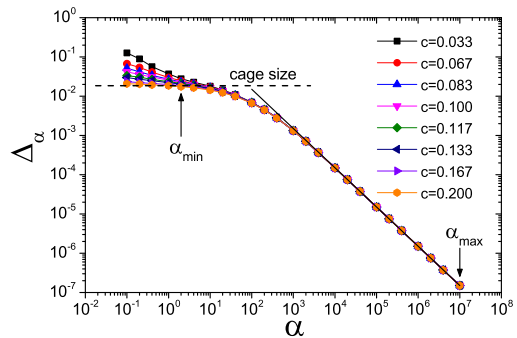}
\caption{The mean-squared displacement $\Delta_{\alpha}$ at $T=0.45$ for $N=1200$ used for the Frenkel-Ladd 
thermodynamic integration. The horizontal dashed line is the plateau height 
for $c=0.2$ shown in Fig.~\ref{fig1:msd_overlap}(a).
The solid line corresponds to the behavior of the Einstein solid, $\Delta_{\alpha}=3/(2\alpha)$.
}
\label{fig2:fl}
\end{figure}

To advance on this point, we have accurately determined $s_{\rm vib}$ by
taking into account the effect of anharmonic vibrations, using two
independent approaches.  The first one makes use of the idea of Frenkel and
Ladd~\cite{frenkel1984new,coluzzi1999thermodynamics,sastry2000evaluation,berthier2017configurational,williams2018experimental}
of introducing a series of systems which parametrically interpolates
between the original system and an Einstein solid, and then to carry
out a thermodynamic integration to calculate the vibrational entropy
of the original system. In practice we have introduced the Hamiltonian
$\beta H(\alpha) = \beta H(0) + \alpha \sum_{i=1}^{(1-c)N}|{\bf r}_i -
{\bf r}_{0 i}|^2$, where $H(0)$ is the original Hamiltonian, $\beta=1/T$,
$\alpha$ is a spring constant, and ${\bf r}_{0 i}$ is the equilibrium
configuration of the particle $i$ in the original system.  We then
measure the MSD $\Delta_{\alpha}=\sum_{i=1}^{(1-c)N} [\langle |{\bf r}_i -
{\bf r}_{0 i}|^2 \rangle]/(1-c)N$ from which we obtain the entropy\\[-8mm]

\begin{equation} 
s_{\rm FL} =  s_{\rm E}(\alpha_{\rm max}) + \int_{0}^{\alpha_{\rm max}} \mathrm{d} \alpha \Delta_{\alpha},
\label{eq1}
\end{equation} 

\noindent
where $s_{\rm E}(\alpha_{\rm max})$ is the entropy of the Einstein solid.
Figure~\ref{fig2:fl} shows $\Delta_{\alpha}$ for $10^{-1} \leq \alpha
\leq 10^7$.  For $\alpha > 10^4$, this function follows very closely
$3/(2 \alpha)$, the behavior of the Einstein solid, and hence we can
replace $\Delta_\alpha$ by this expression if $\alpha$ is large. The
entropy for this Einstein solid is then given by $s_{\rm E}(\alpha_{\rm
max})=\frac{3}{2} - 3 \ln \Lambda -\frac{3}{2} \ln \left(\frac{\alpha_{\rm
max}}{\pi}\right)$, where $\Lambda$ is the de Broglie thermal wavelength.
In practice we have set $\alpha_{\rm max} = 10^7$.

\begin{figure}[tb]
\includegraphics[width=0.90\columnwidth]{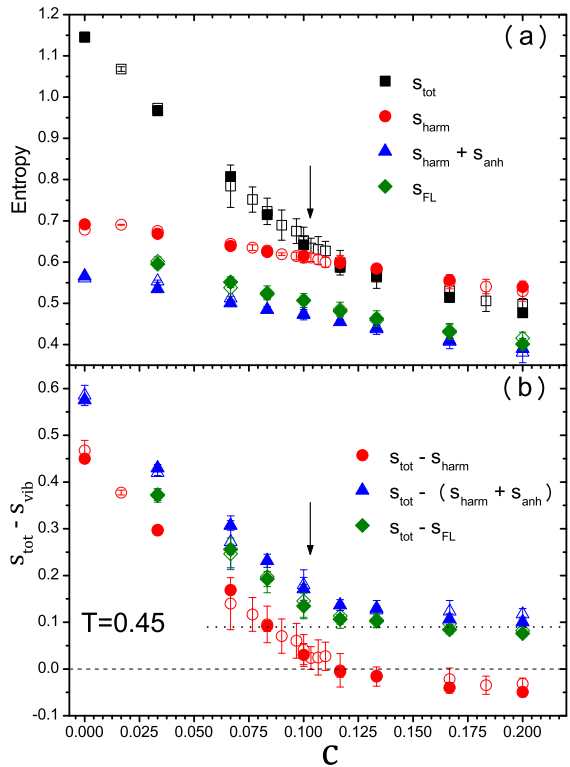}
\caption{(a) $c-$dependence of the total entropy $s_{\rm tot}$, as well
as different estimates of the vibrational entropy. (b) $c-$dependence
of $s_{\rm tot}-s_{\rm vib}$. The open and filled symbols are for 
$N=300$ and $N=1200$, respectively.
The horizontal arrows locate $T_{\rm K}(c)$ where the skewness of the overlap distribution becomes zero~\cite{ozawa2015equilibrium}.}
\label{fig3:entropy}
\end{figure}  

At small $\alpha$, the dependence of $\Delta_{\alpha}$ on $\alpha$ becomes
weak suggesting that in this parameter range the particles are vibrating
in a cage created by the {\it original} Hamiltonian and not by the one of the
Einstein solid. However, at the smallest $\alpha$, $\Delta_{\alpha}$ is
not completely flat, since at long times the particles are able to escape
slowly from their cages. The height of the plateau in $\Delta_{\alpha}$
allows to estimate the amplitude of the vibrations in the real system,
i.e., to determine the harmonic and anharmonic component of the motion.
Indeed, the height of the plateau in $\Delta_{\alpha}$ is consistent with
that of $\Delta(t)$ from Fig.~\ref{fig1:msd_overlap}(a). To estimate the
entropy that is due to the vibrational motion we can evaluate the integral
given by Eq.~(\ref{eq1}) by replacing the contribution to the integral
for $\alpha < \alpha_{\rm min}=2$ by $\alpha_{\rm min} \Delta_{\alpha_{\rm
min}}$, thus removing in this manner the contribution of the relaxational
part to the entropy. The so obtained vibrational entropy $s_{\rm FL}$
is shown in Fig.~\ref{fig3:entropy}(a) (green diamonds). As expected,
the $c-$dependence of $s_{\rm FL}$ is smooth and shows no apparent
singularity. 

\begin{figure*}[tb]
\includegraphics[width=0.6\columnwidth]{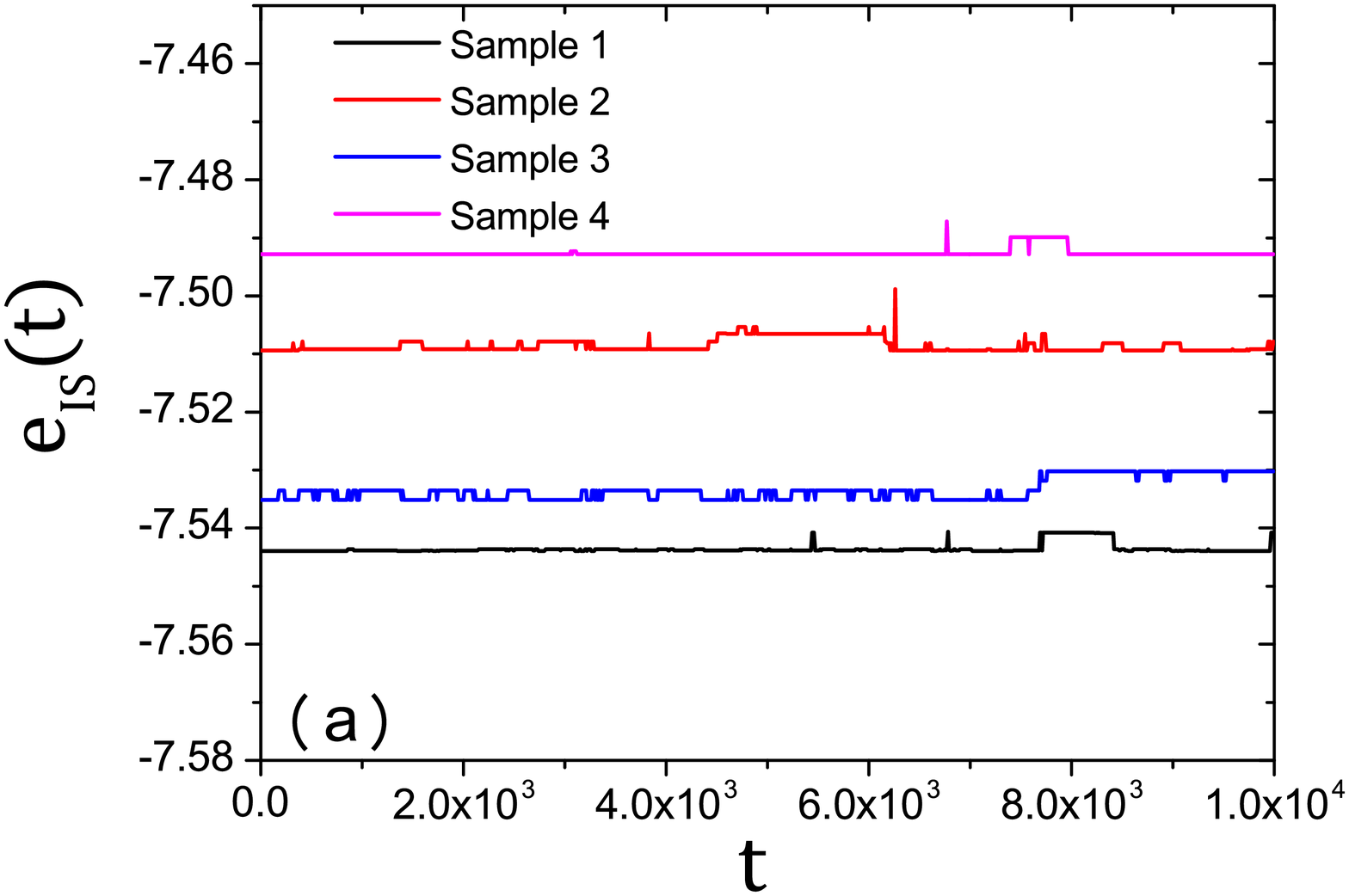}
\includegraphics[width=0.6\columnwidth]{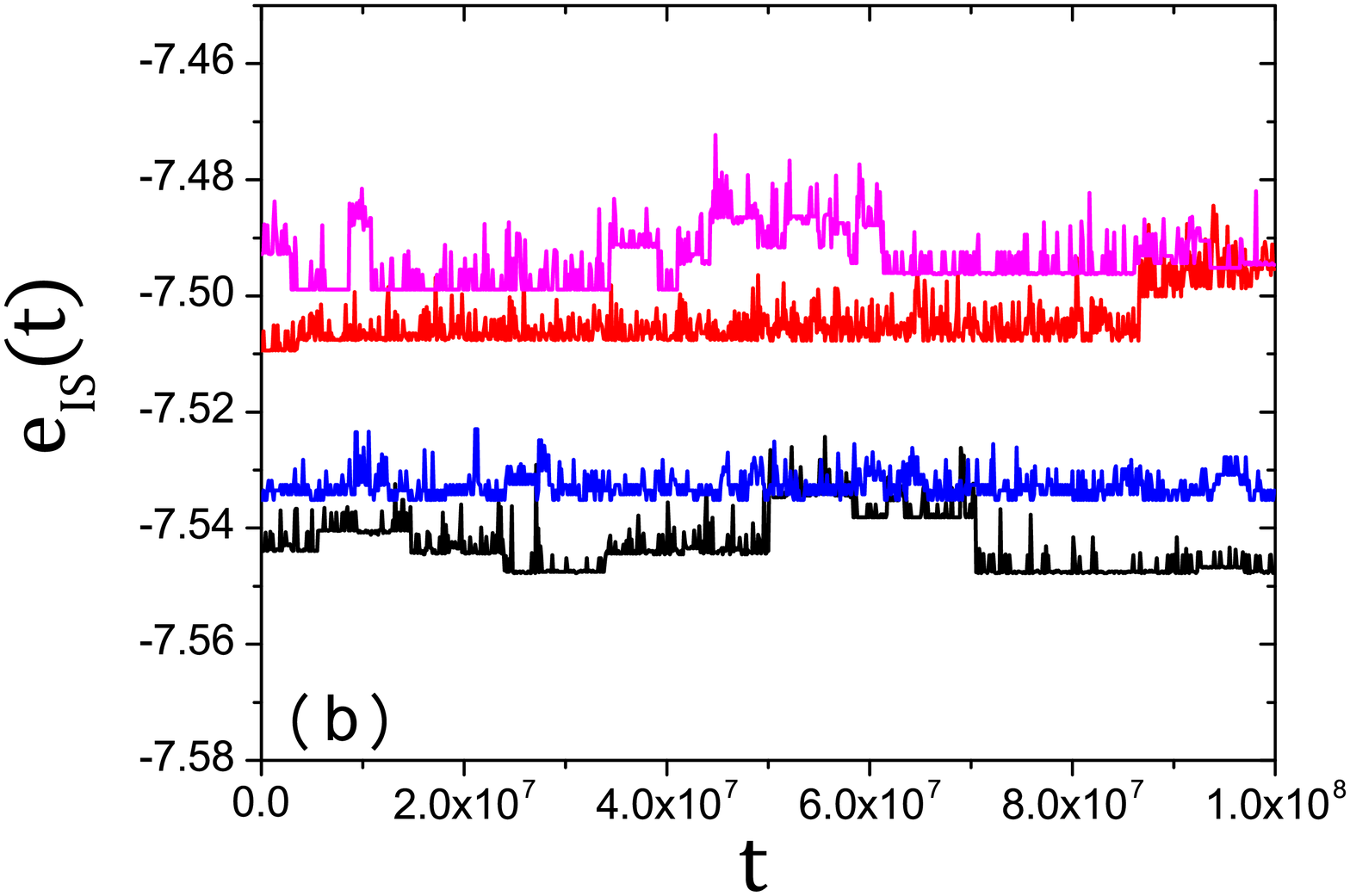}
\includegraphics[width=0.6\columnwidth]{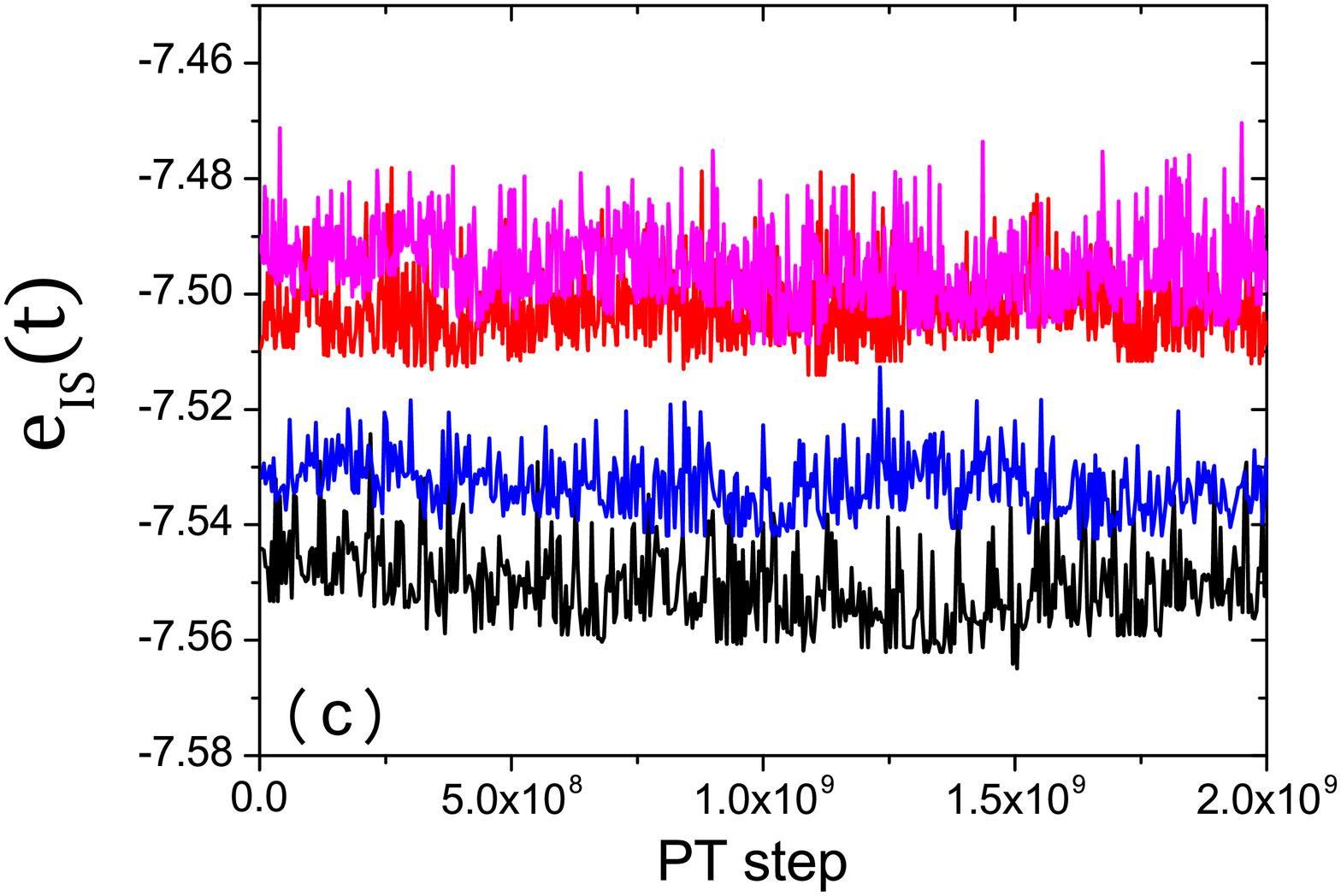}
\vspace*{-0mm}

\includegraphics[width=0.5\columnwidth,trim={0 0 0 30mm},clip]{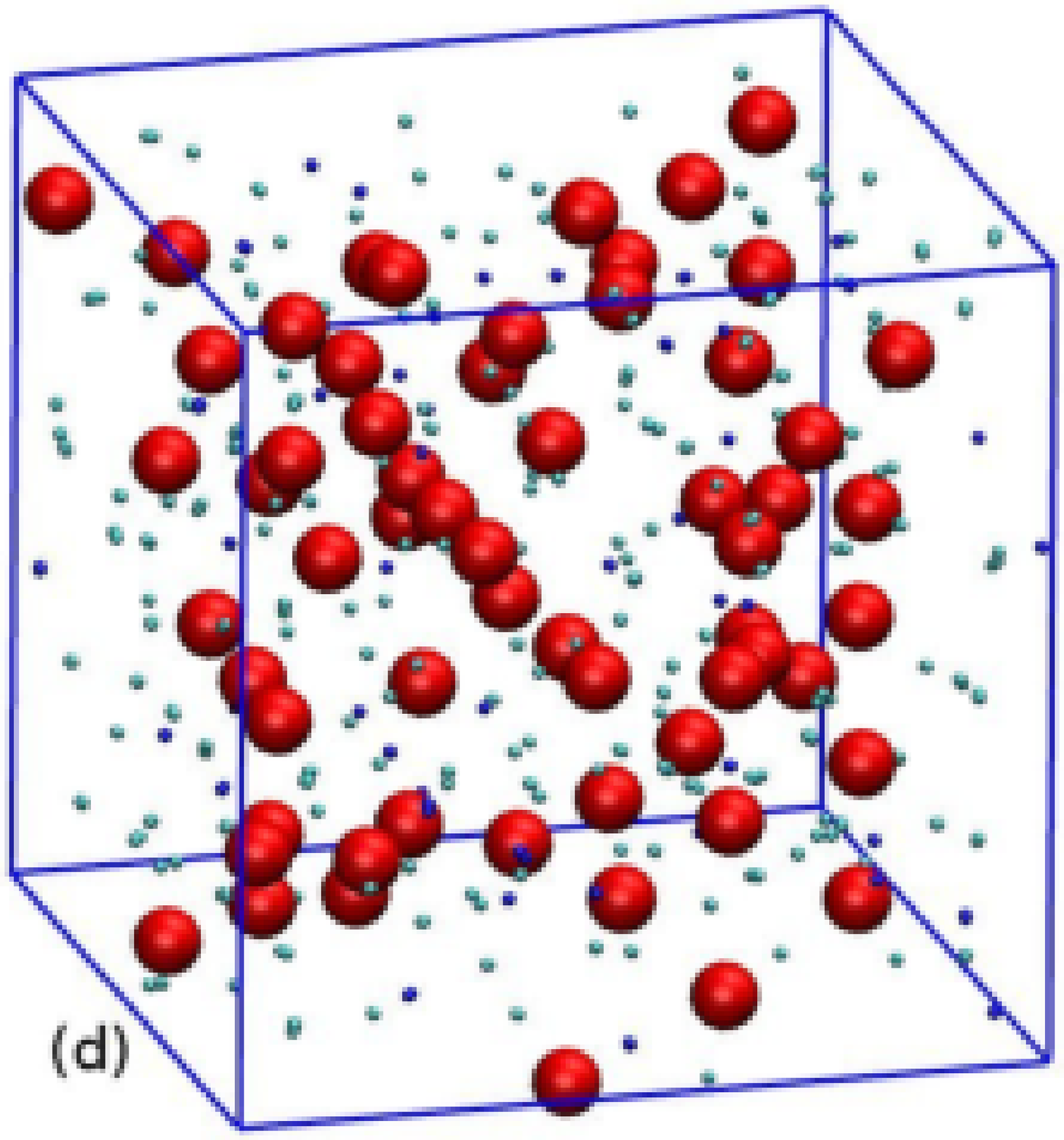}
\includegraphics[width=0.5\columnwidth,trim={0 0 0 30mm},clip]{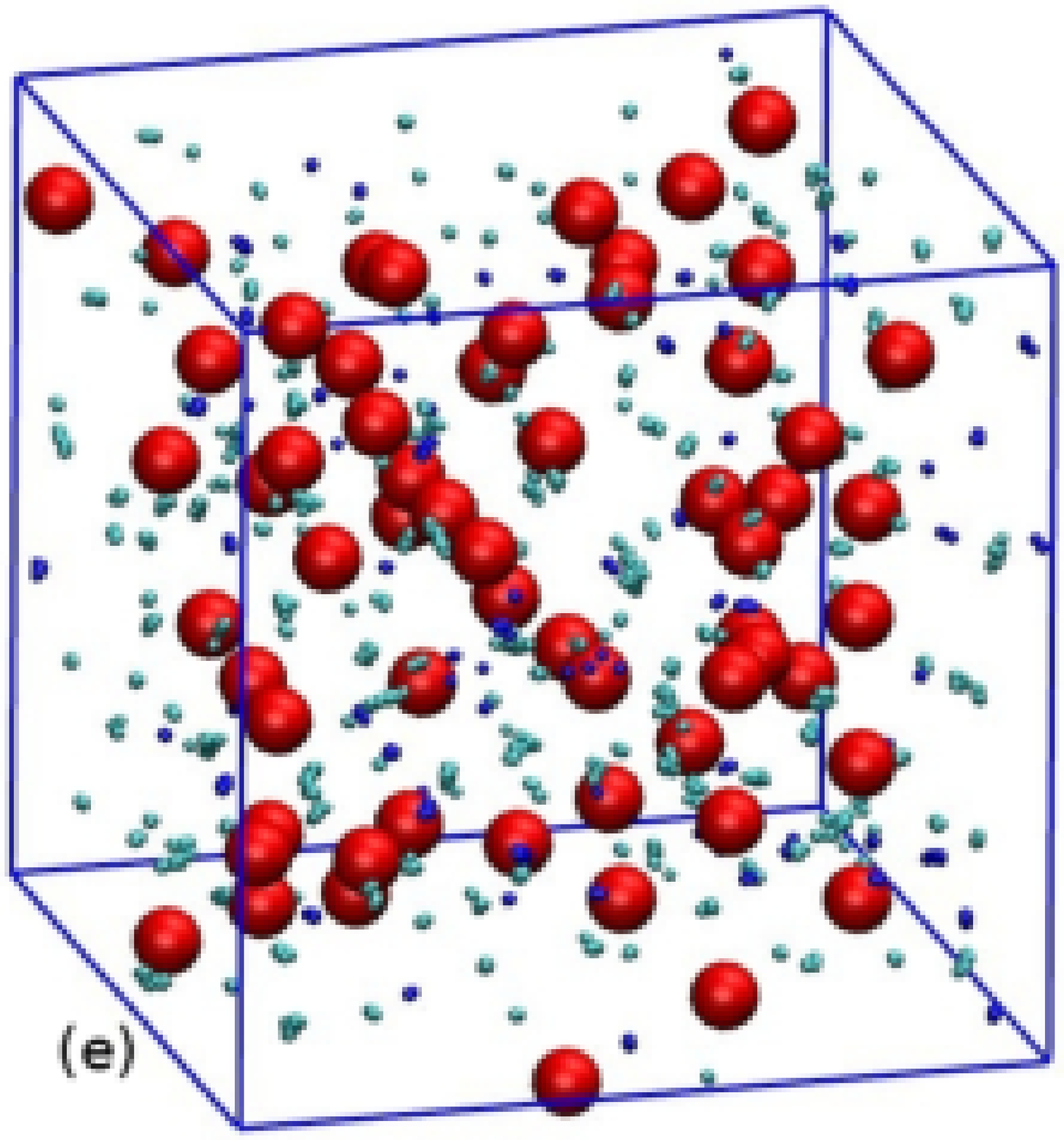}
\includegraphics[width=0.5\columnwidth,trim={0 0 0 30mm},clip]{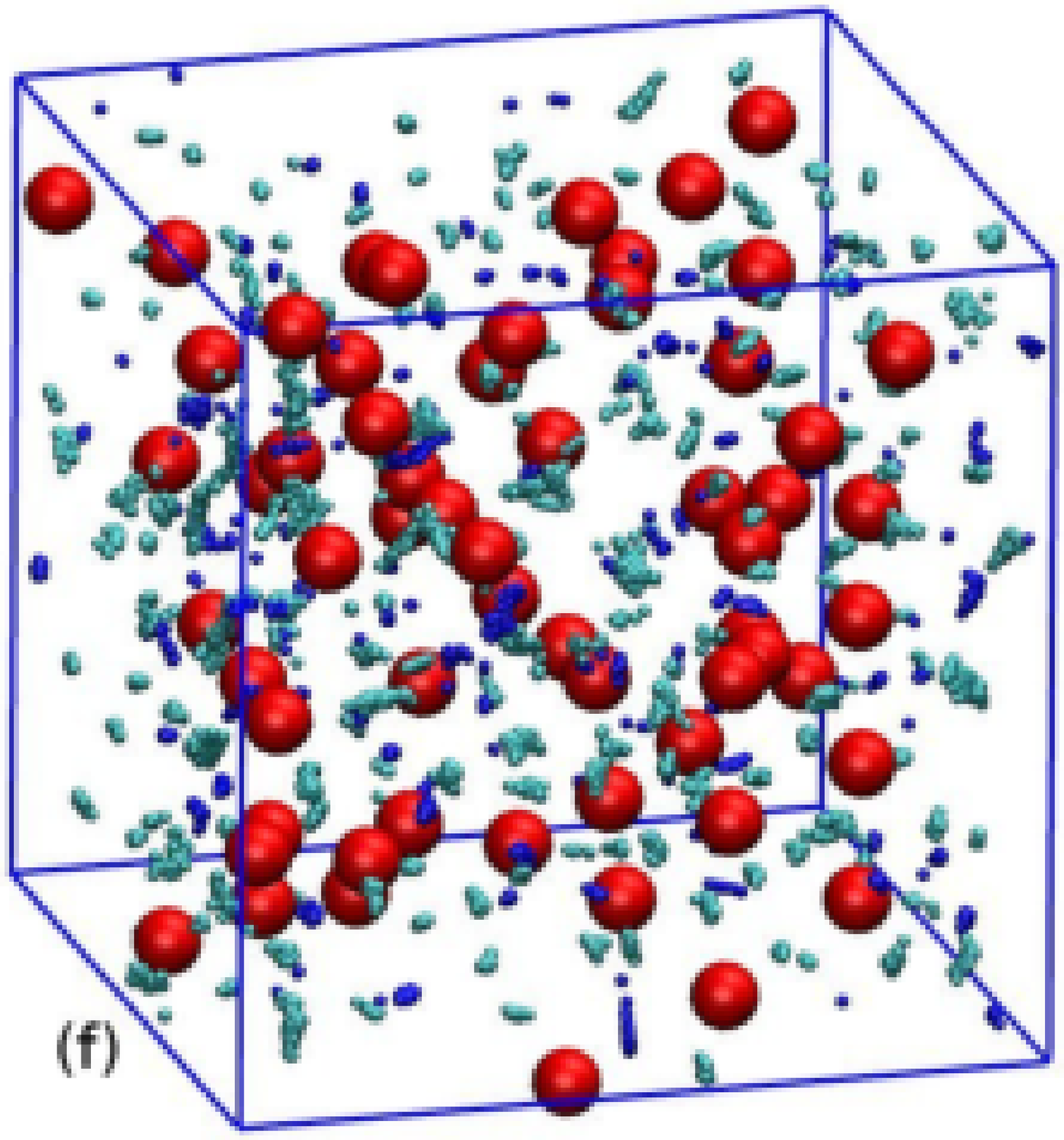}
\vspace*{-8mm}

\caption{(a-c): $e_{\rm IS}(t)$ at $T=0.45$ and $c=0.2$ for $N=300$ samples with (a) $10^4$ and
(b) $10^8$ MC time step, and (c) $2 \times 10^9$ PT time step.  Different lines correspond to different samples. (d-f):
Superposition of the corresponding snapshots (sample 1) of the IS configurations with (d) $10^4$ and
(e) $10^8$ MC time step, and (f) $2 \times 10^9$ PT time step. Pinned particles are shown in red, and mobile A and B particles
are shown in blue and gray, respectively.
}
\label{fig:IS}
\end{figure*}

The second approach to estimate the anharmonic contribution to the
vibrational entropy is to determine the difference between the potential
energy of the system and its inherent structure energy and then to
subtract the harmonic part, giving thus the anharmonic part of the
vibrational energy~\cite{mossa2002dynamics,sciortino2005potential}.
One finds that the so obtained energy is {\it negative} which makes
that the resulting value for the vibrational entropy, $s_{\rm harm}
+ s_{\rm anh}$, is smaller than $s_{\rm harm}$ (see blue triangles in
Fig.~\ref{fig3:entropy}(a)), a result that agrees with previous
studies~\cite{mossa2002dynamics,sciortino2005potential}. We see that
this quantity agrees very well with our estimate for the vibrational
entropy as obtained from the Frenkel-Ladd procedure, indicating that we
have determined it with good precision.

Figure~\ref{fig3:entropy}(b) shows the $c-$dependence of 
$\Delta s = s_{\rm tot}-s_{\rm vib}$. We see that the improved estimate
for $s_{\rm vib}$ makes that now $\Delta s$ no longer goes to zero
even at large $c$. Instead it shows a kink at the concentration at
which the order parameter had a jump~\cite{ozawa2015equilibrium},
indicating that at this point the thermodynamic properties of the system
have a singular behavior, i.e., that the phase diagram determined in
Ref.~\cite{ozawa2015equilibrium} is not altered by this improved estimate
of $s_{\rm vib}$. Note that we show data for the two system sizes and
within the accuracy of the data we see no finite size effects.

To determine the origin of the finite value of $\Delta s$
in the ideal glass phase we probe the potential energy landscape of the
system~\cite{goldstein1969viscous,stillinger1982hidden,wales2003energy,sciortino2005potential,heuer2008exploring}.
Since at low $T$ the configuration space can be decomposed into
the basins of attraction of the inherent structures and vibrations
around the IS, an investigation of the PEL should help to understand
the nature of the motion of the system.  Figure~\ref{fig:IS}(a)
shows the time dependence of the IS energy $e_{\rm IS}(t)$,
\cite{schroder2000crossover,denny2003trap}, for $10^4$ steps of MC
dynamics at $c=0.2$, a timescale that corresponds to the vibrations
inside the cage (see Fig.~\ref{fig1:msd_overlap}(a)). We see that $e_{\rm
IS}(t)$ remains basically constant but shows some spikes indicating
that the system accesses for a short period an excited state before it
falls back to the original IS, implying that on this time scale only a
single IS is relevant~\cite{heuer2008exploring}. Figure~\ref{fig:IS}(d)
shows the superposition of the configurations for these IS's and we
see that basically all of them are identical, in agreement with the
result from Fig.~\ref{fig1:msd_overlap}(a) that on this time scale the
correlation function does not decay. (For the sake of comparison we show
in the SI a corresponding plot for the fluid state.) If the time window
is increased by a factor of $10^4$, Fig.~\ref{fig:IS}(b), we find that
$e_{\rm IS}(t)$ starts to make larger excursions and that these are no
longer reversible, indicating that the system explores new IS. This
is the reason why on this time scale the correlation function decay
slightly, see Fig.~\ref{fig1:msd_overlap}(b). Figure~\ref{fig:IS}(e)
shows that now there are indeed many different IS's, but since the
corresponding positions of the particles form small clusters we can
conclude that the configurations are quite similar. Finally we show in
Fig.~\ref{fig:IS}(c) the evolution of $e_{\rm IS}(t)$ as obtained from a
PT run. We see that in this case the value of $e_{\rm IS}(t)$ fluctuates
quite significantly, but that these fluctuations are still smaller than
the sample to sample fluctuations and therefore we can infer that there
are indeed many different IS's even in the equilibrium glass state. This
conclusion is corroborated by the real space image of the configurations,
Fig.~\ref{fig:IS}(f), which now shows quite a few small clusters, i.e.,
the positions of the particles in the different IS's differ by a small
amount. The presence of these different IS's is thus the reason for the
finite value of $\Delta s$. We emphasize that the clusters seen in
Fig.~\ref{fig:IS}(f) do not depend on the length of the PT run.

Our results show that the dynamic and thermodynamic properties of a system
with pinned particles give a coherent picture of the equilibrium
glass state. Although quantities like the self intermediate scattering
function~\cite{chakrabarty2015dynamics,chakrabarty2016understanding,ozawa2015reply}
or the self van Hove function (see SI) do decay to zero, the
collective functions do not, once the critical pinning concentration
is surpassed.  At this transition point the order parameter shows
a jump~\cite{ozawa2015equilibrium} and the entropy a kink.
Our finding that even in the equilibrium glass state particles are able
to explore more than one IS indicates that care has to be taken in
the definition of the configurational entropy: $s_{\rm conf}$ should
not be identified as the difference between the total entropy and the
purely vibrational part, since such a definition gives rise to a non-zero
$s_{\rm conf}$ even in the ideal glass state. Instead $s_{\rm conf}$ is related
to the number of local free energy minima~\cite{biroli2000inherent} and
for the case of the pinned system such a minimum is a {\it collection}
of IS that are geometrically close together. The motion of the particles
inside this local free energy minimum is the reason for the {\it partial}
relaxation of the collective correlation functions. Although the present
study concerns a pinned system, it can be expected that bulk systems have
a similar behavior since local inhomogeneities in the structure
will make that certain regions in the very deeply supercooled liquid are
already frozen in, thus leading to a very high value of the viscosity,
whereas other regions are still mobile. So the evoked problem with the
correct definition of the configurational entropy is likely to exist
also in the case of bulk systems.

\begin{acknowledgments} We thank G. Biroli, C. Dasgupta, K. Hukushima,
Y. Jin, S. Karmakar, and K. Kim for helpful discussions. The numerical simulations were
partially performed at Research Center of Computational Science (RCCS), Okazaki, Japan. W.~K. acknowledges ANR-15-CE30-0003-02 , and A.~I. and
K. M acknowledge JSPS KAKENHI Grants Number JP16H04034, JP17H04853,
JP25103005, and JP25000002.
\end{acknowledgments}

\vspace*{20mm}
\appendix
{\bf \Large Supplemental Information}

\section{Dynamics}

\subsection{Absence of aging}

\begin{figure}[tb]
\includegraphics[width=0.8\columnwidth]{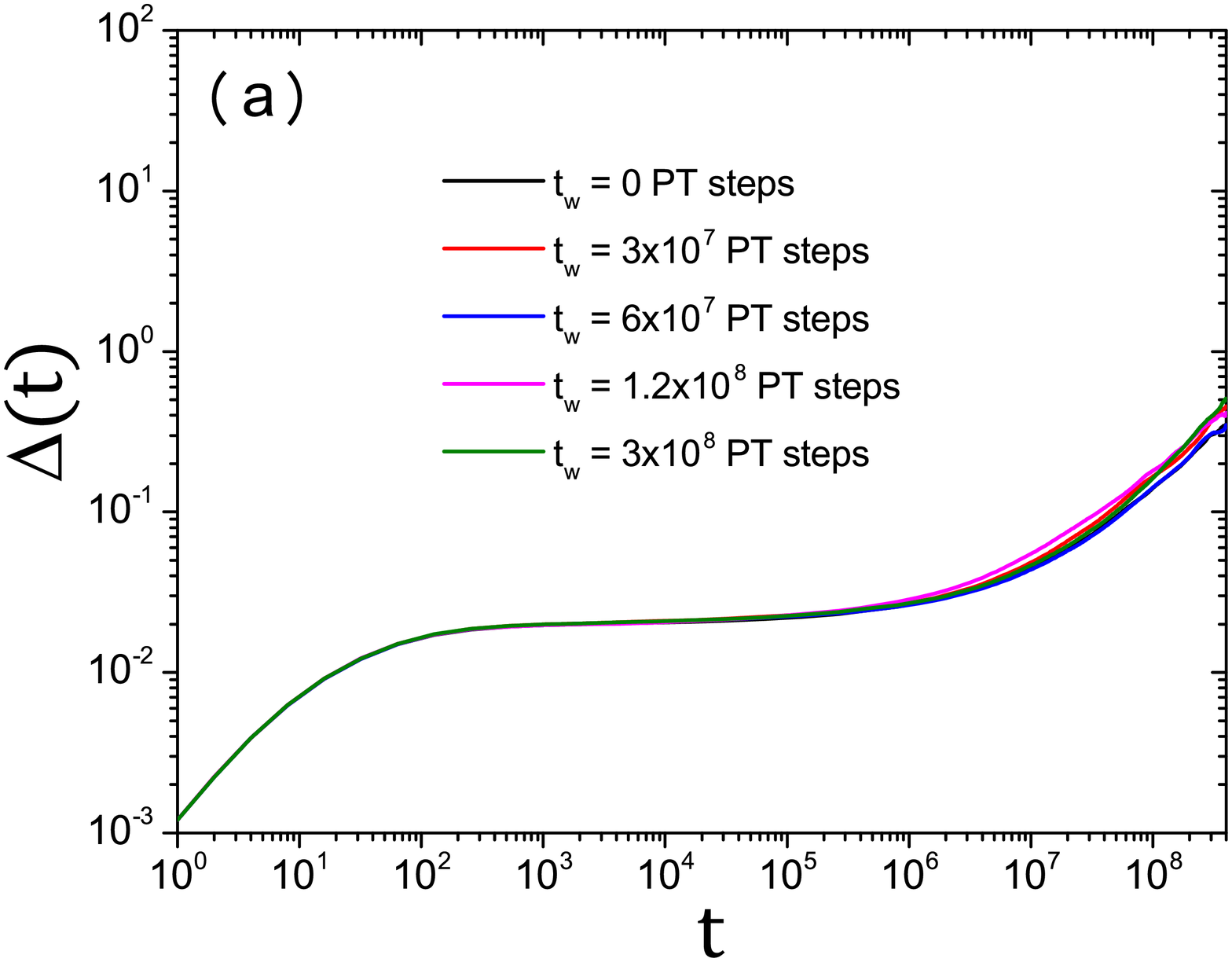}
\includegraphics[width=0.8\columnwidth]{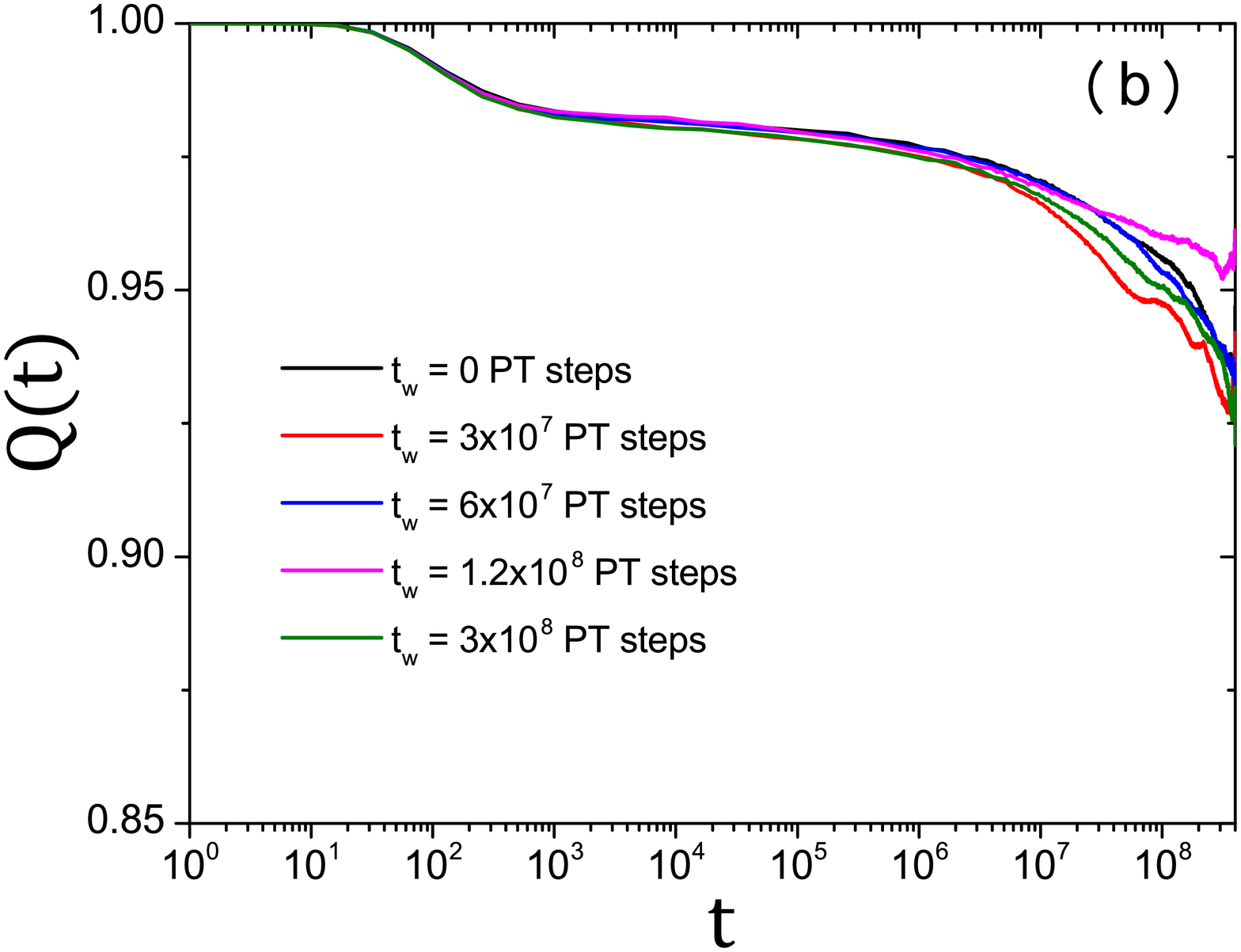}
\caption{
(a): Time dependence of the mean squared displacement $\Delta(t)$ of
a single sample at $T=0.45$ and $c=0.167$ for $N=1200$.  The starting
configurations correspond to different times $t_{\rm w}$ in the PT
trajectory and are given in the legend. (b) Same as in panel (a) but now
for the collective overlap function $Q(t)$ defined in the main text.  }
\label{fig:aging_check}
\end{figure}

In Fig.~1(a) of the main text we show that the mean squared displacement
(MSD) $\Delta(t)$ of a tagged particle shows at intermediate times
the usual plateau found in glassy systems~\cite{binder2011glassy}
but that, surprisingly, the MSD turns at long times upward even when
the system is in the ideal glass phase. In order to demonstrate that
this upturn is not related to any kind of non-equilibrium dynamics, see
e.g. Ref~\cite{kob1997aging}, we show in Fig.~\ref{fig:aging_check}(a)
the MSD as obtained for different origins of time. In practice we have
taken the trajectory as obtained from the parallel tempering (PT) run
and selected five different configurations from that trajectory. For
each one of these configurations we have started a standard MC run and
have obtained in this manner five different MSDs.  The figure shows that
these curves are identical with high precision, i.e., they do not depend
on the starting configurations. This is hence strong evidence that the
upturn of $\Delta(t)$ seen at large times is not related to aging
dynamics since in that case one finds that the time for the
upturn shifts to increasingly longer times if the age of the system,
i.e., the time at which the initial configuration for the MC run was
selected, is increased~\cite{kob1997aging}. The same conclusion is
reached if one considers the time dependence of the overlap, again
calculated starting from different starting configurations, see
Fig~\ref{fig:aging_check}(b).

\subsection{Dynamics as characterized by the van Hove function}

In Fig.~1 of the main text we show the time-dependence of the self
dynamics (i.e., the mean squared displacement) as well as a collective
correlation function, i.e., the collective overlap function $Q(t)$. In
Ref.~\cite{ozawa2015reply} we have argued that because of the presence
of the pinned particles the self correlation functions behave very
differently from the collective ones in that the latter will decay to
a plateau with finite value whereas the former will go to zero as in a
normal liquid. To understand better the motion of the particles in real
space it is useful to look at the self and distinct parts of the van Hove correlation
functions~\cite{hansen2006theory}.

\begin{figure*}[htb]
\includegraphics[width=0.66\columnwidth]{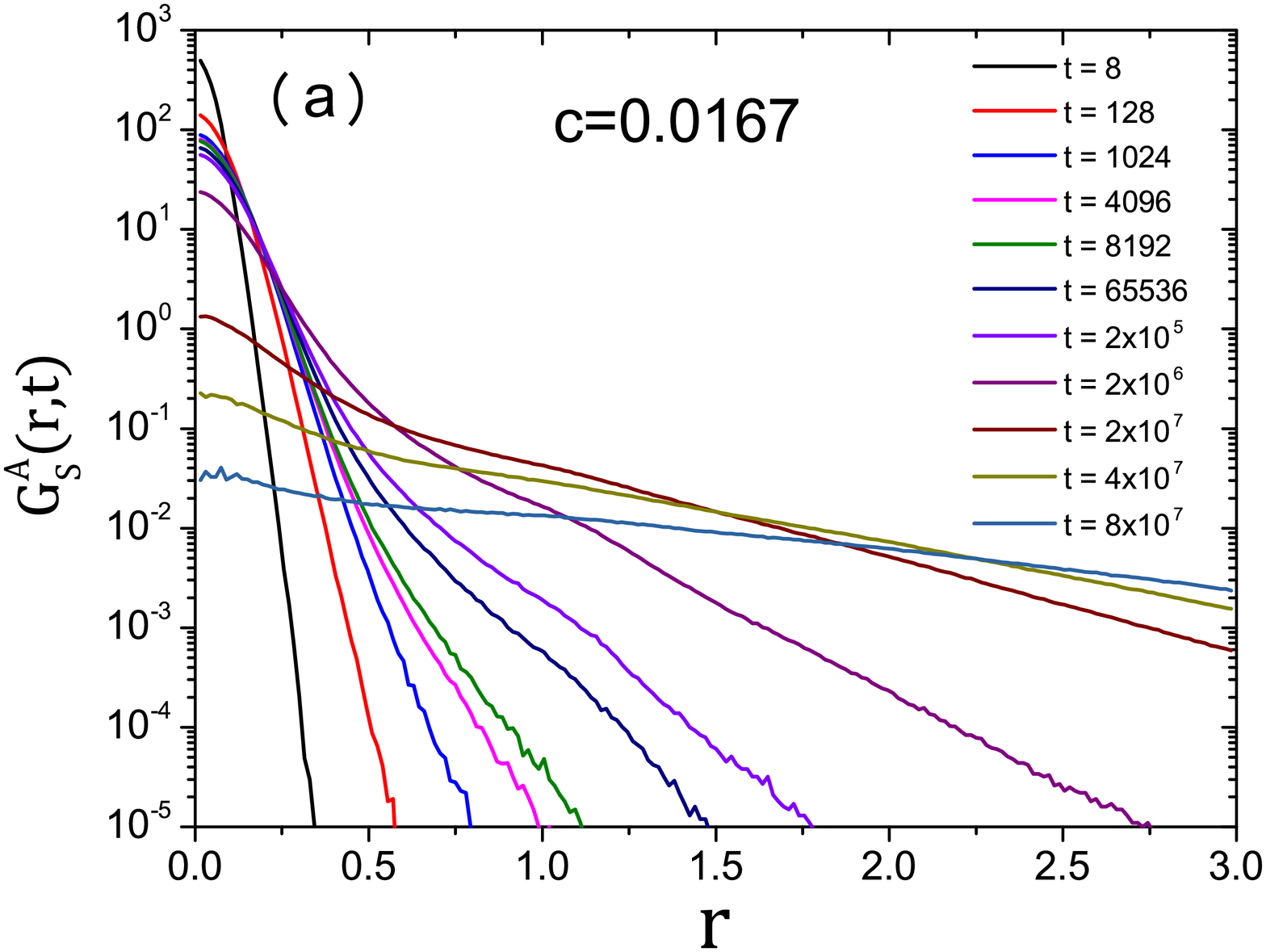}
\includegraphics[width=0.66\columnwidth]{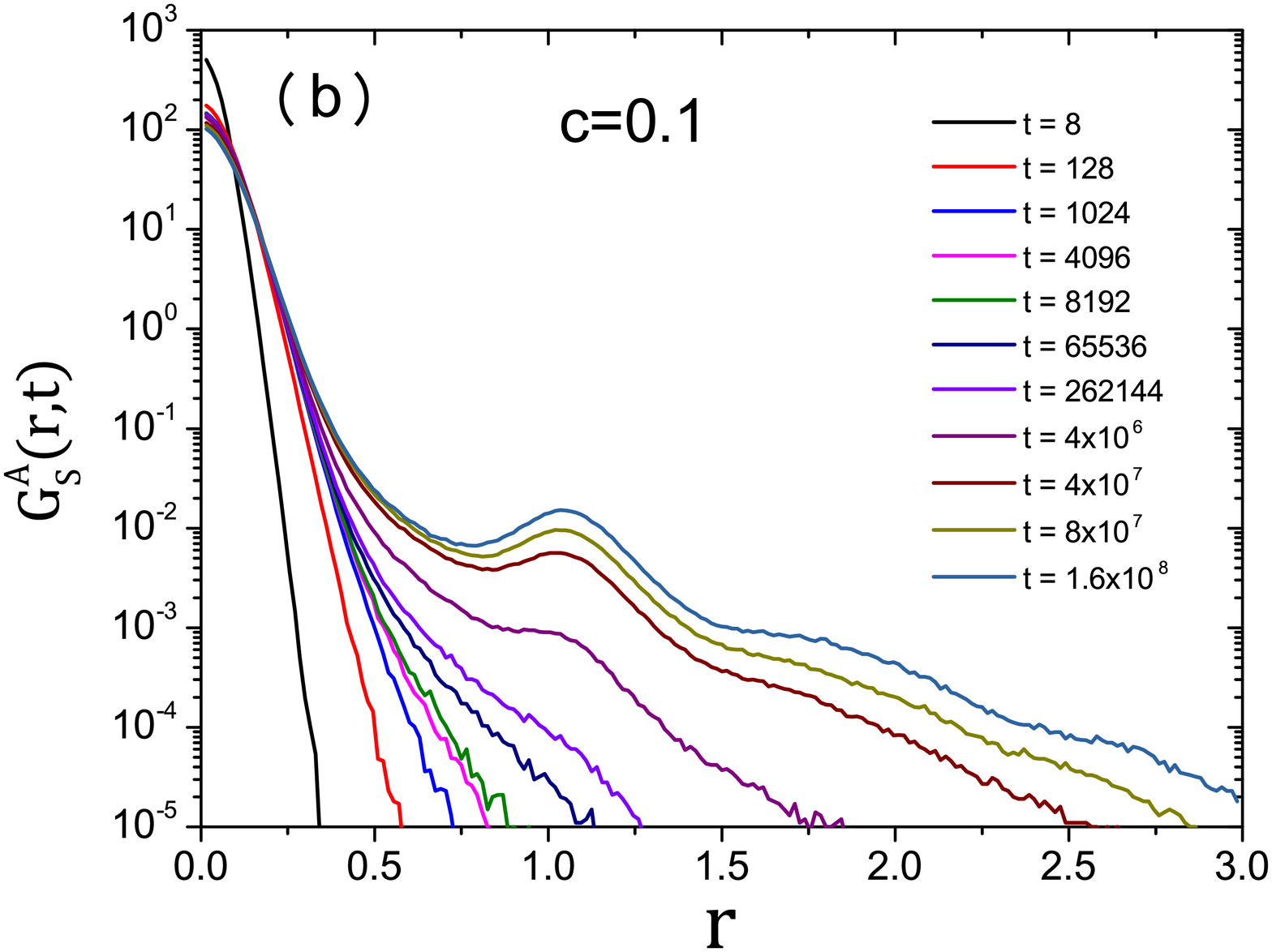}
\includegraphics[width=0.66\columnwidth]{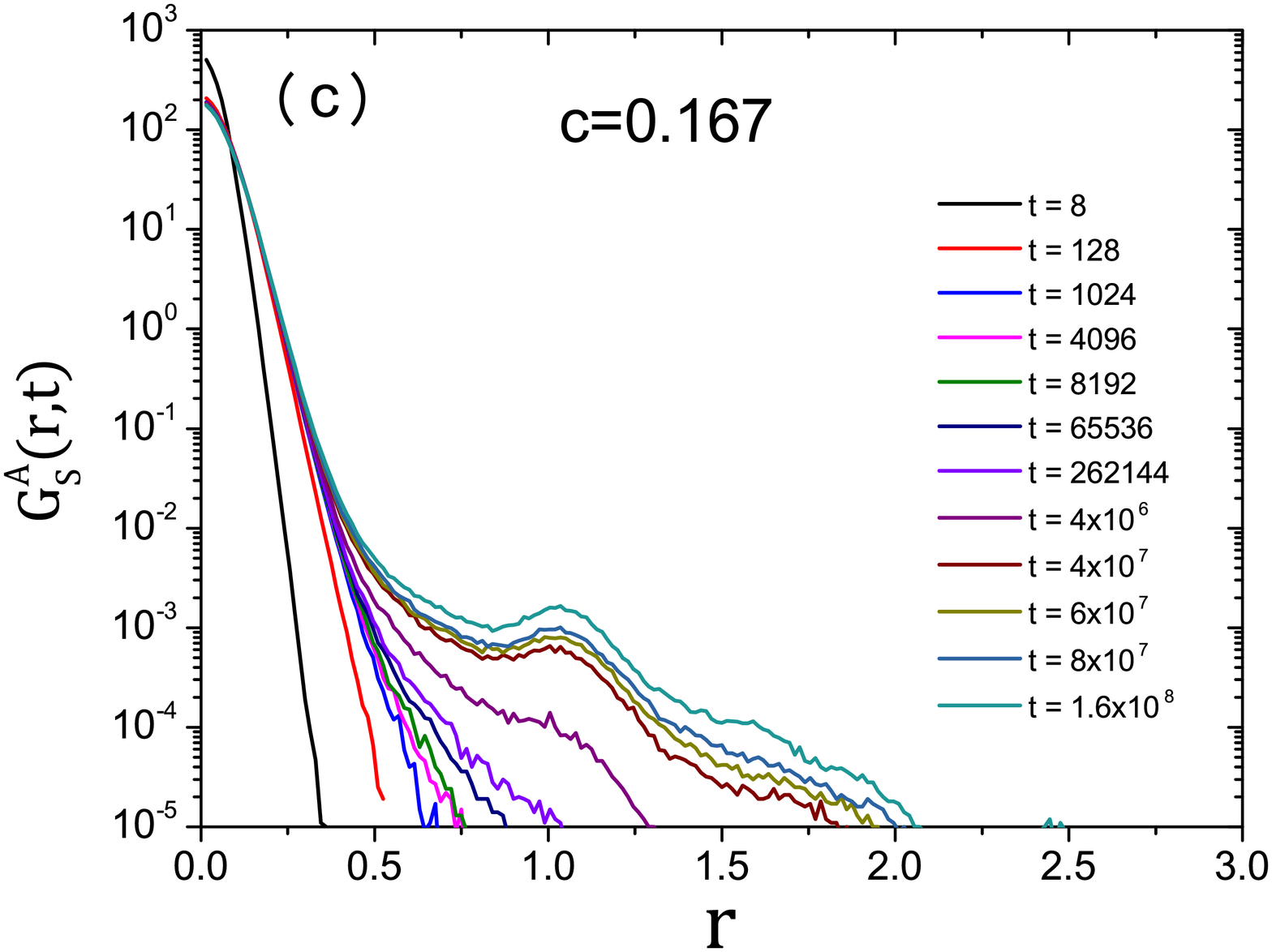}
\includegraphics[width=0.66\columnwidth]{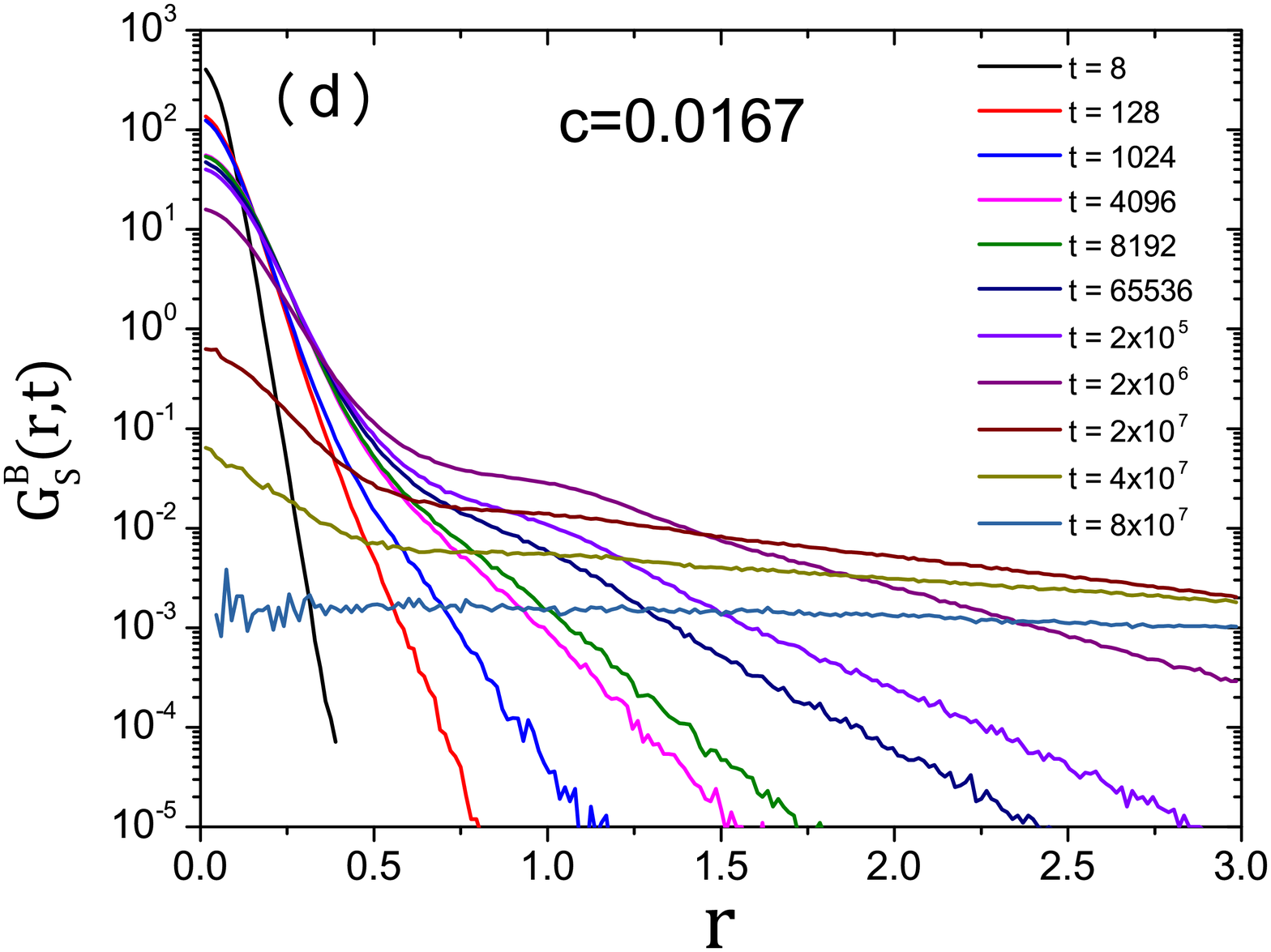}
\includegraphics[width=0.66\columnwidth]{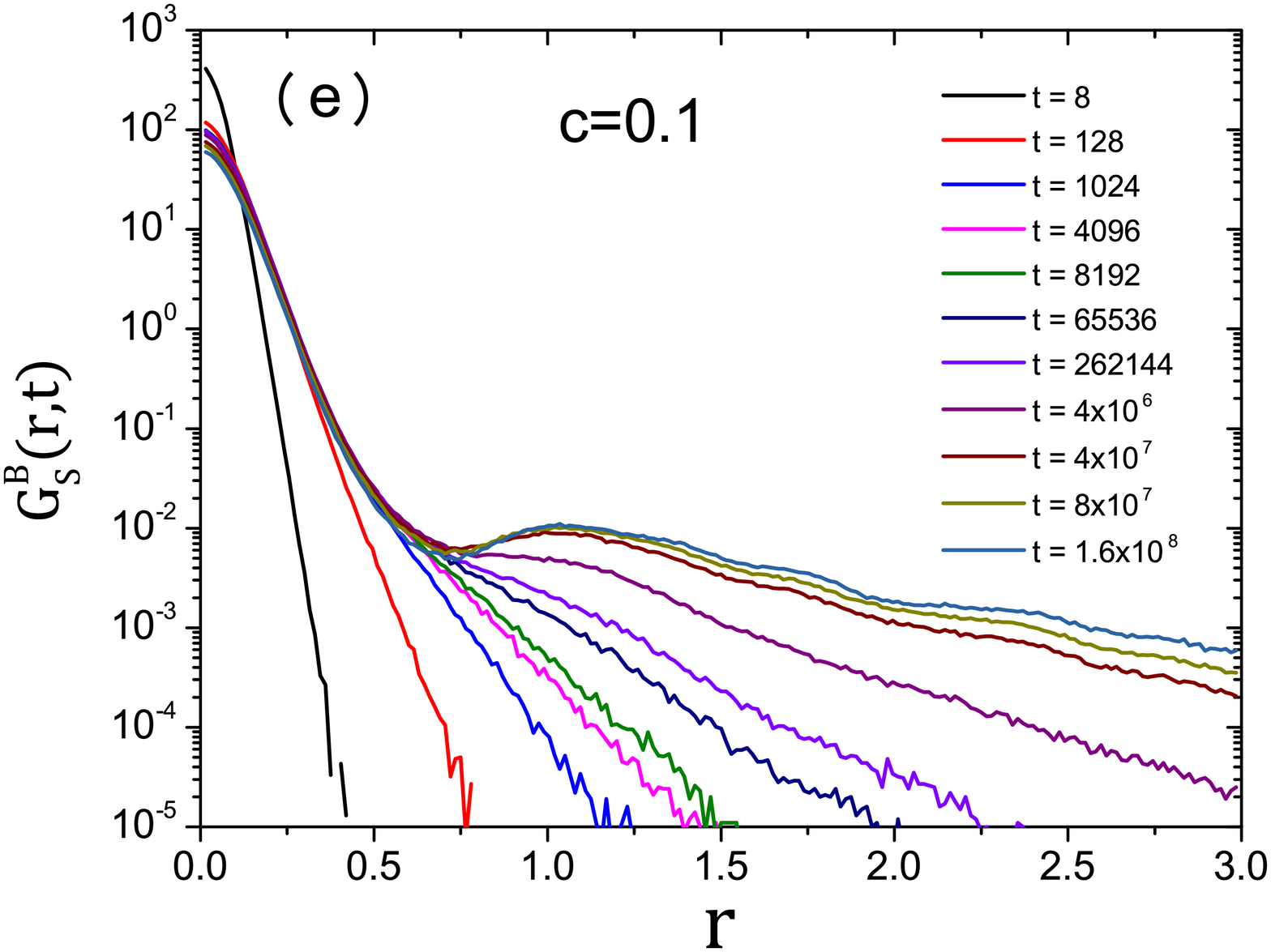}
\includegraphics[width=0.66\columnwidth]{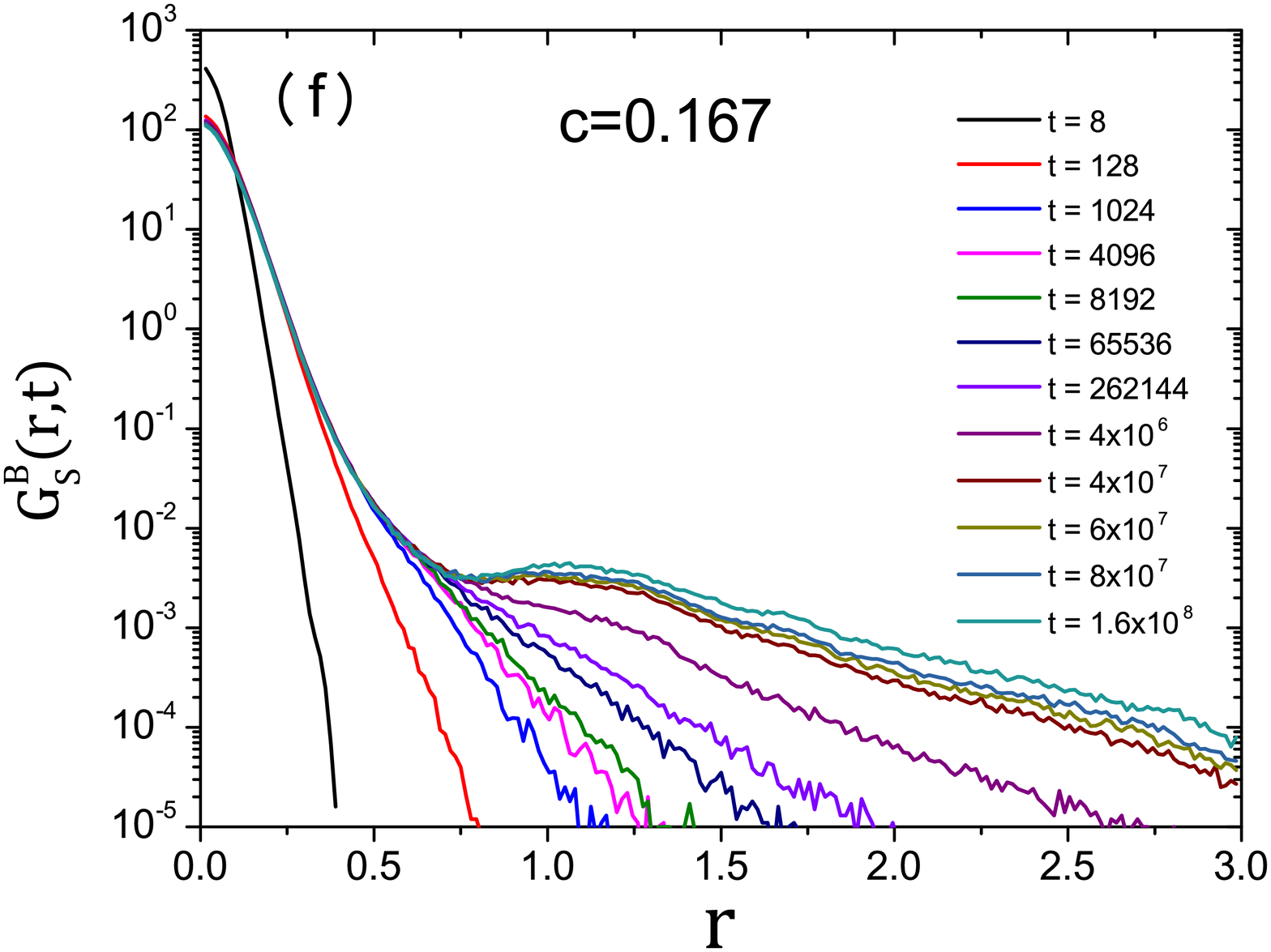}
\caption{
Self part of the van Hove correlation function at $T=0.45$ and different values of $c$ for $N=1200$.
(a)-(c): Species A, (d)-(f): Species B.
}
\label{fig:vanHove_self}
\end{figure*}  

The self part of the van Hove function is defined by
\begin{equation}
G_{\rm s}^{\alpha}(r,t)= \frac{1}{(1-c)N_{\alpha}} \sum_{i=1}^{(1-c)N_{\alpha}} 
\left[ \left\langle  \delta(r-|{\bf r}_i(0)-{\bf r}_i(t)|) \right\rangle \right],
\label{eq_SI_1}
\end{equation}

\noindent
where $N_\alpha$ is the number of particles of species $\alpha$.
Here $\alpha, \beta \in \{ {\rm A}, {\rm B}\}$.
The distinct part of the van Hove function is defined by

\begin{widetext}
\begin{eqnarray}
G_{\rm d}^{\alpha \alpha}(r,t) &=& \frac{N}{(1-c)N_{\alpha}^2} 
\sum_{i=1}^{(1-c)N_{\alpha}} \sum_{j (\neq i)}^{(1-c)N_{\beta}} \left[ \left\langle 
\delta(r-|{\bf r}_i(0)-{\bf r}_j(t)|) \right\rangle \right], \\
G_{\rm d}^{\alpha \beta}(r,t) &=& \frac{N}{(1-c)N_{\alpha} N_{\beta}} 
\sum_{i=1}^{(1-c)N_{\alpha}} \sum_{j}^{(1-c)N_{\beta}} \left[ \left\langle 
\delta(r-|{\bf r}_i(0)-{\bf r}_j(t)|) \right\rangle \right] \quad (\alpha \neq \beta).
\label{eq_SI_2}
\end{eqnarray}
\end{widetext}

The $r$-dependence of $G_{\rm s}(r,t)$ is shown in
Fig.~\ref{fig:vanHove_self} for temperature $T=0.45$ and different
pinning concentrations $c$. For $c=0.0167$, i.e., when the system is very
similar to a glassy bulk system, the $G_{\rm s}(r,t)$ shows at very short
times the Gaussian behavior expected for a harmonic system since on this
time scale the particle explores its local cage. With increasing
time the particle leaves that cage and accesses larger distances before
becoming diffusive. If the concentration is increased to
$c=0.1$, Fig.~\ref{fig:vanHove_self}(b), we see that at large times
the distribution function has a marked peak at $r\approx 1$, i.e.,
the particle hops by a nearest neighbor distance.
Small peaks are seen at larger distances at $r\approx 2$ and 2.7, which correspond to
the second and third nearest neighbor distances~\cite{li2015decoupling}.
If $c$ is increased even more, Fig.~\ref{fig:vanHove_self}(c), the
form of $G_{\rm s}(r,t)$ is not altered appreciably, although the probability
at large distances decreases by more than one order of magnitude.  Thus we
can conclude that the presence of the pinned particles change the
nature of the motion from a flow-like motion at low $c$ to a hopping
motion at intermediate and high $c$.

The self part of the van Hove functions discussed so far were for the A particles which
are larger and thus less mobile than the B particles. In Figs.~\ref{fig:vanHove_self} (d)-(f)
we show the same quantity for the B particles. In agreement with
the results for the bulk system~\cite{kob1995testing1}, we see that already
for very low concentration there is a small peak at $r\approx 1$,
i.e., the particles show some tendency for a hopping dynamics. If $c$ is
increased the peak becomes more pronounced but remains smaller
than the corresponding peak for the A particles.

\begin{figure*}[htb]
\includegraphics[width=0.66\columnwidth]{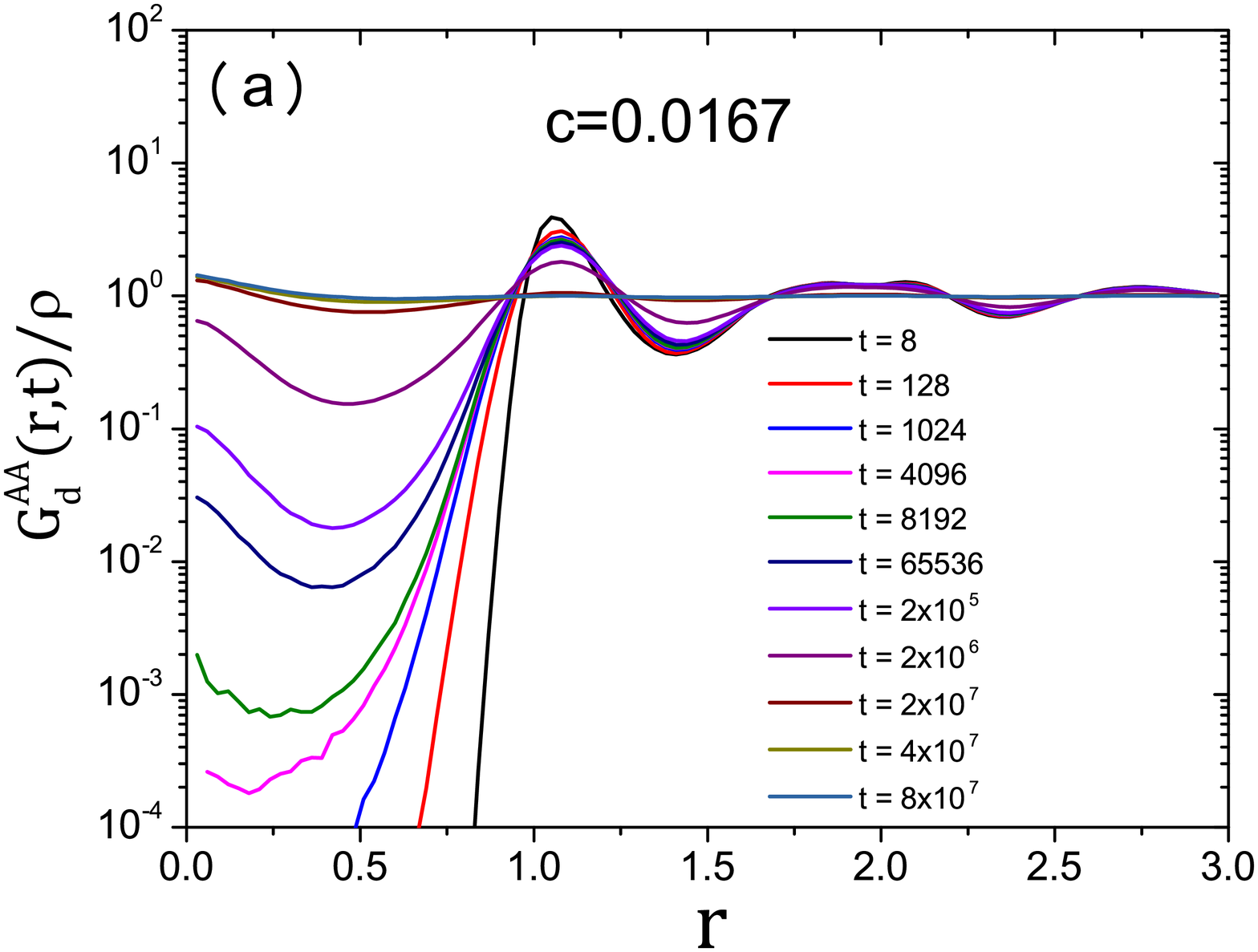}
\includegraphics[width=0.66\columnwidth]{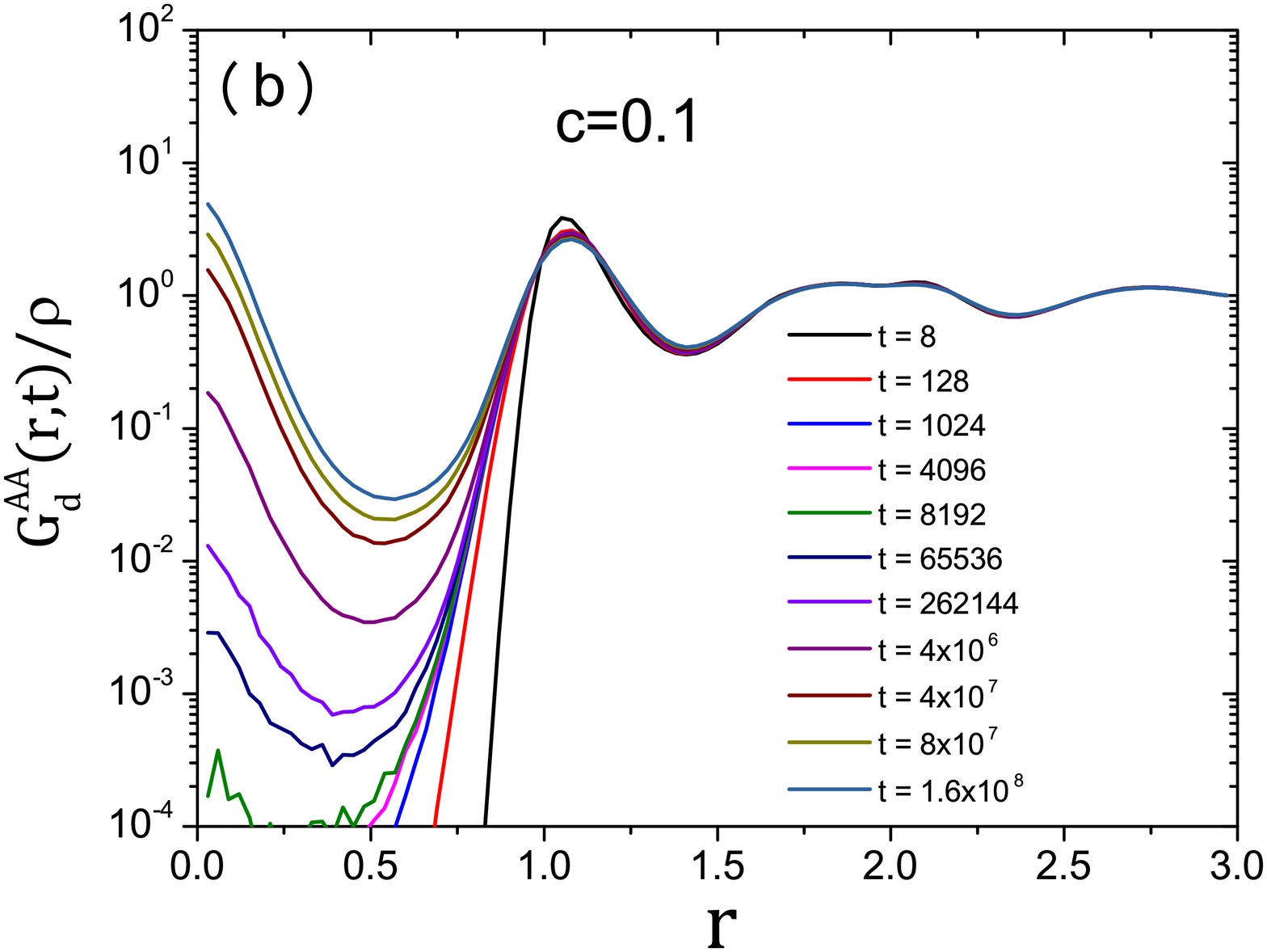}
\includegraphics[width=0.66\columnwidth]{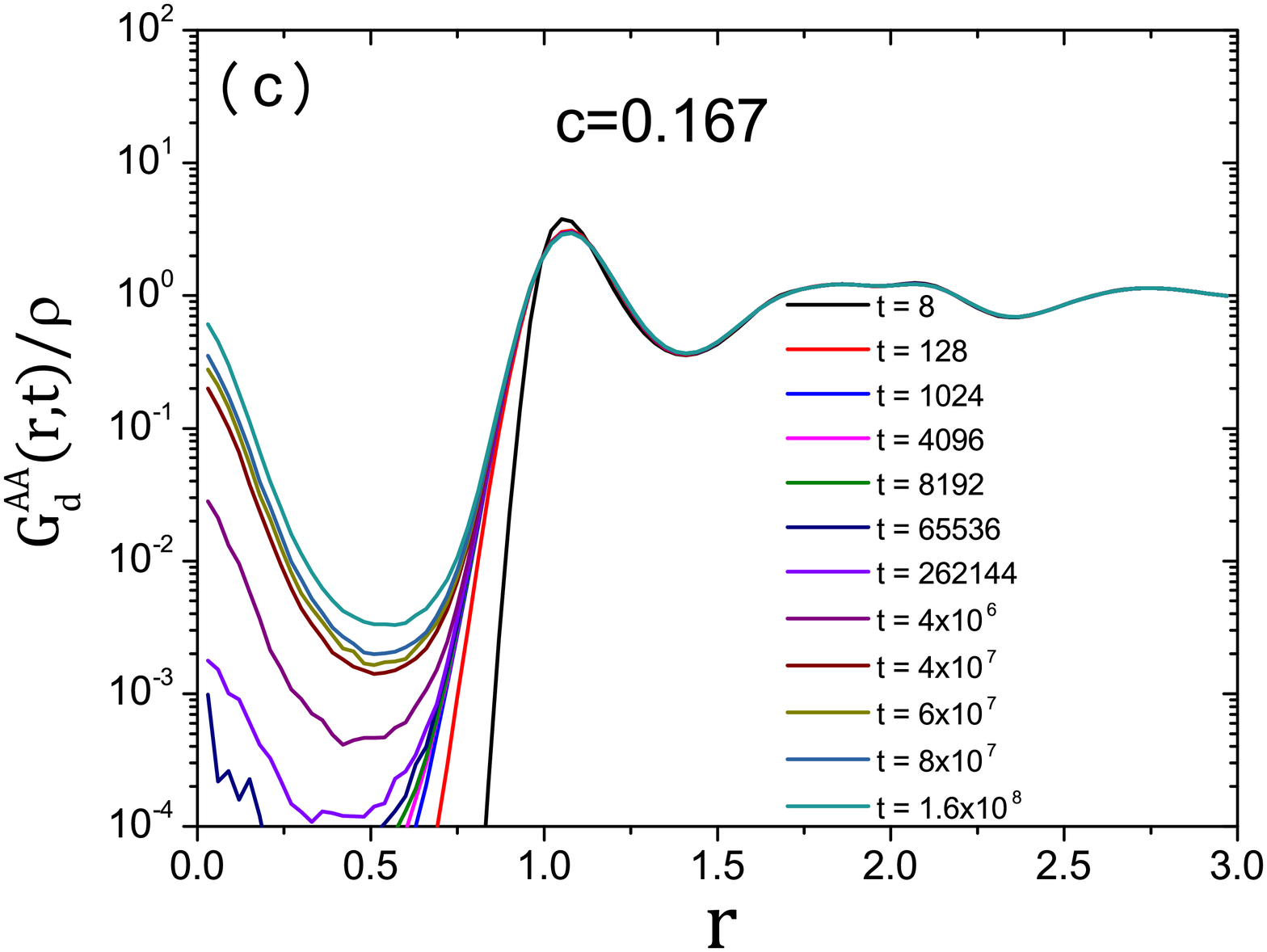}
\includegraphics[width=0.66\columnwidth]{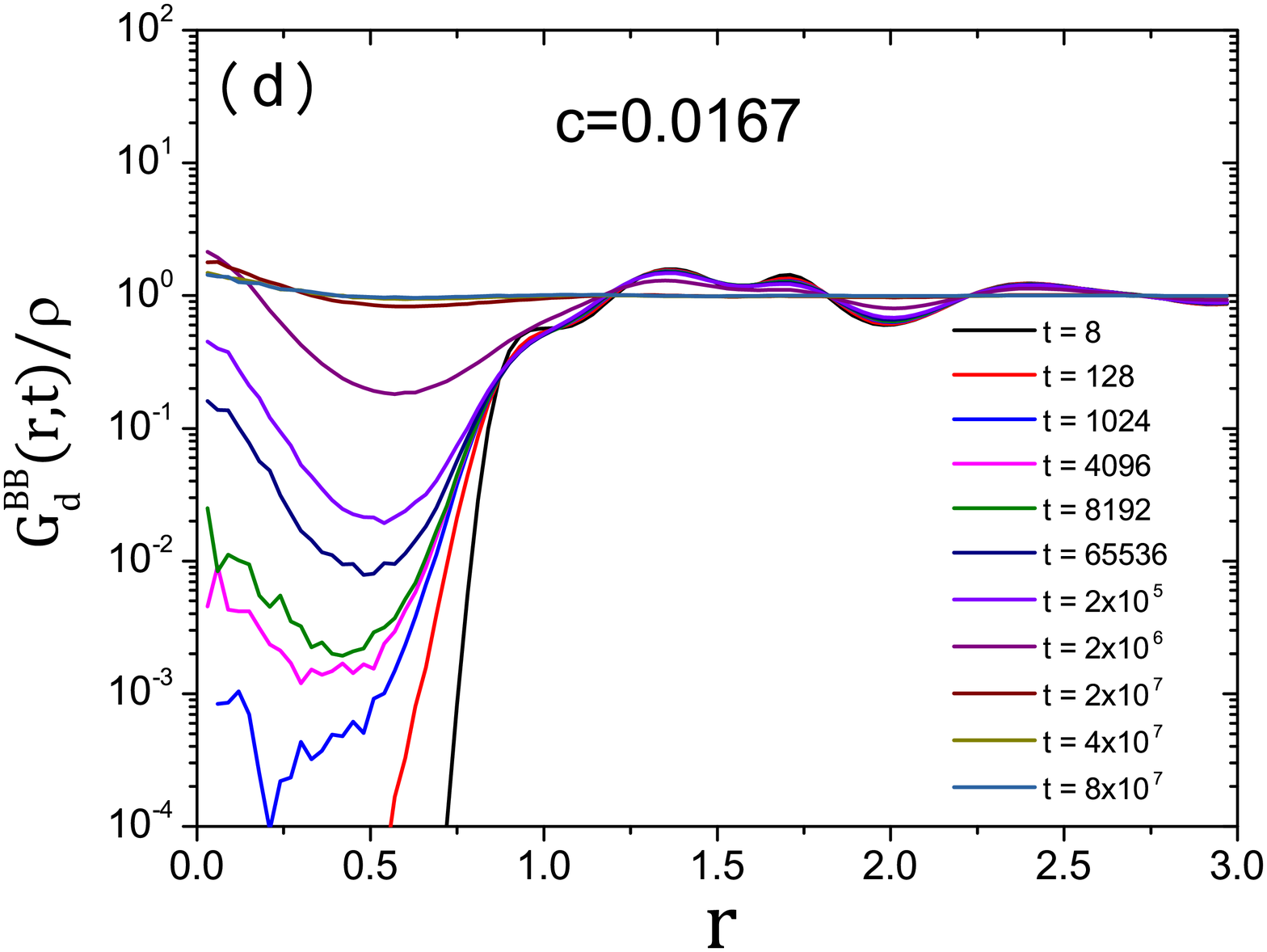}
\includegraphics[width=0.66\columnwidth]{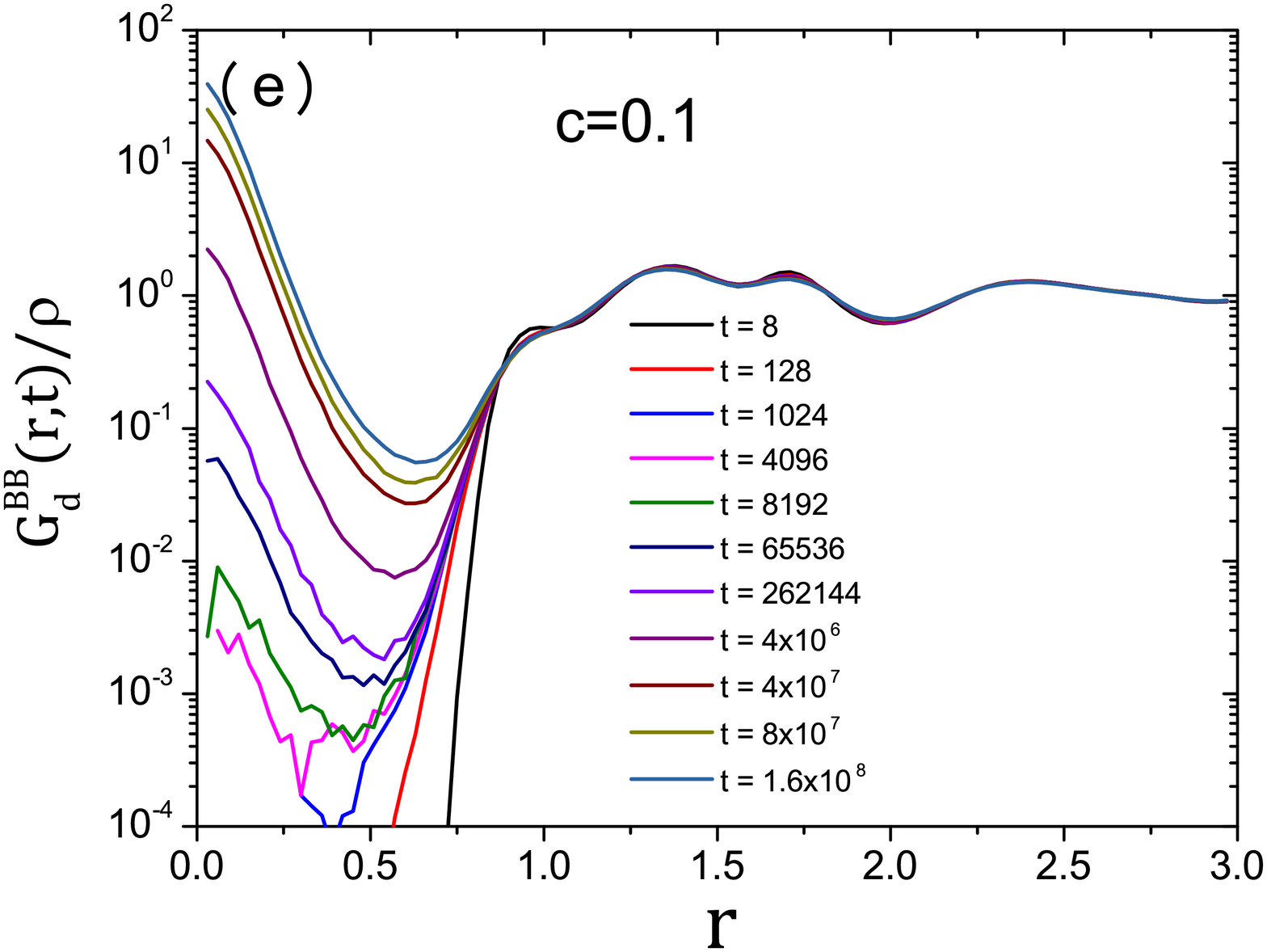}
\includegraphics[width=0.66\columnwidth]{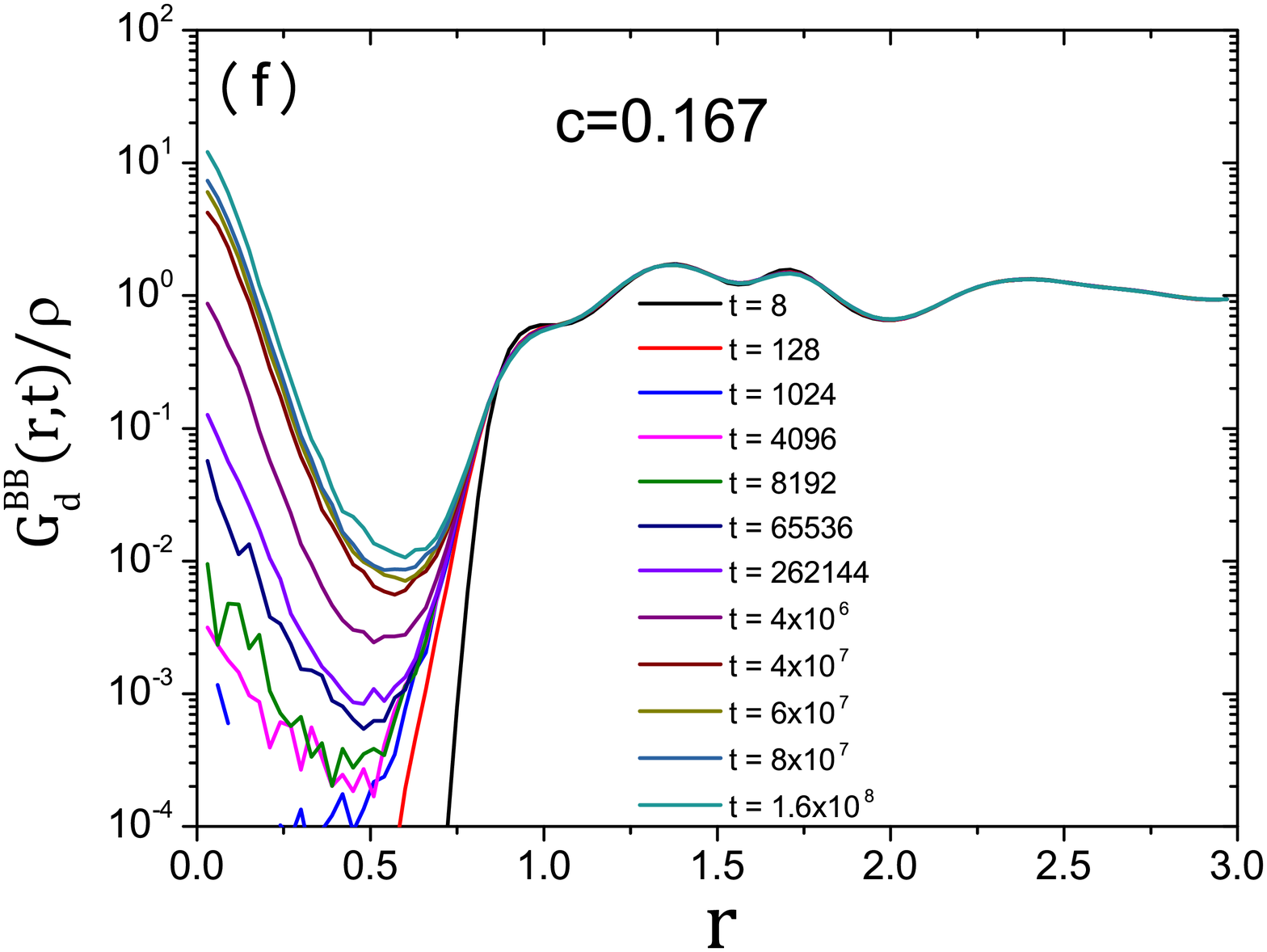}
\includegraphics[width=0.66\columnwidth]{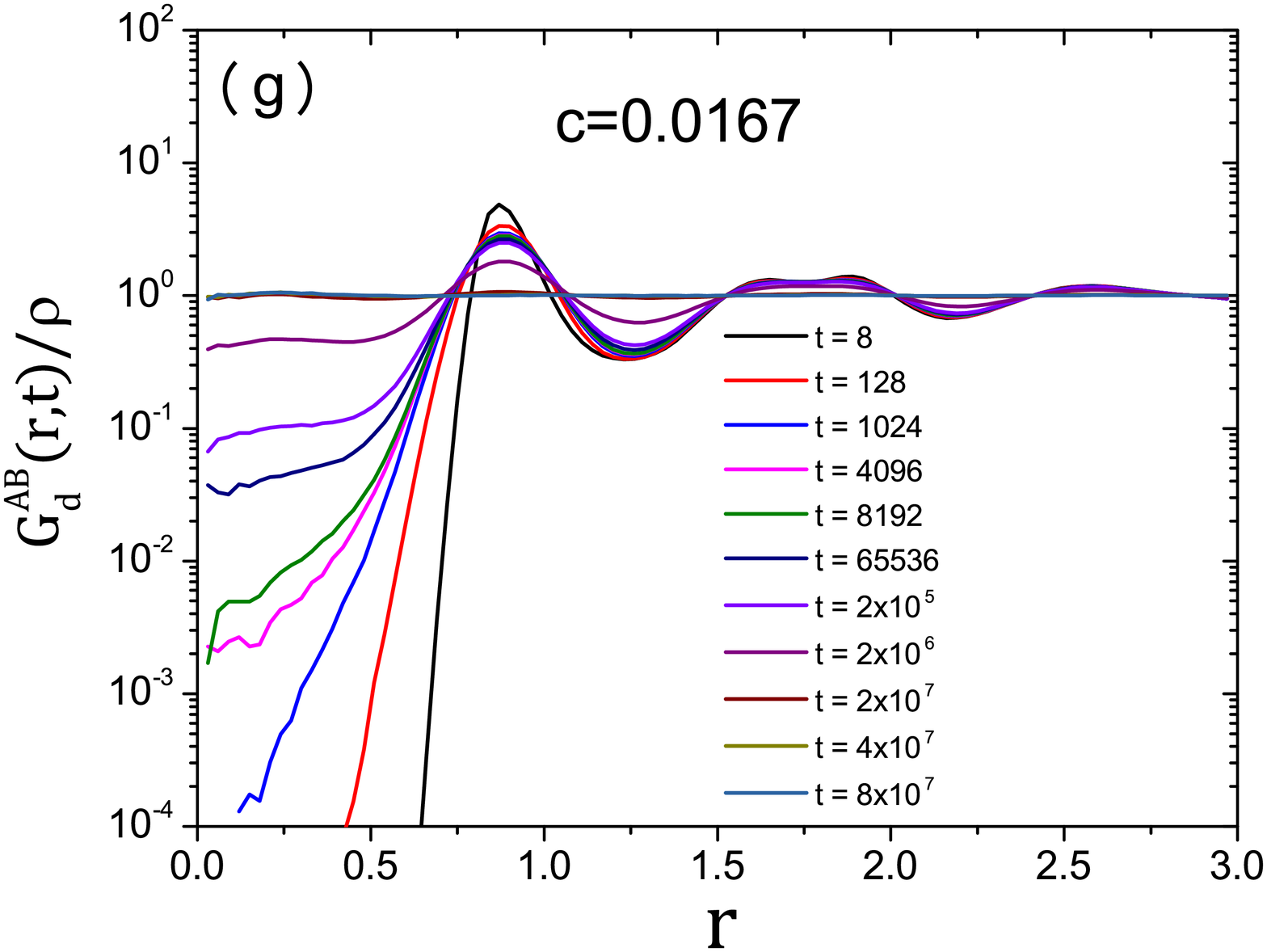}
\includegraphics[width=0.66\columnwidth]{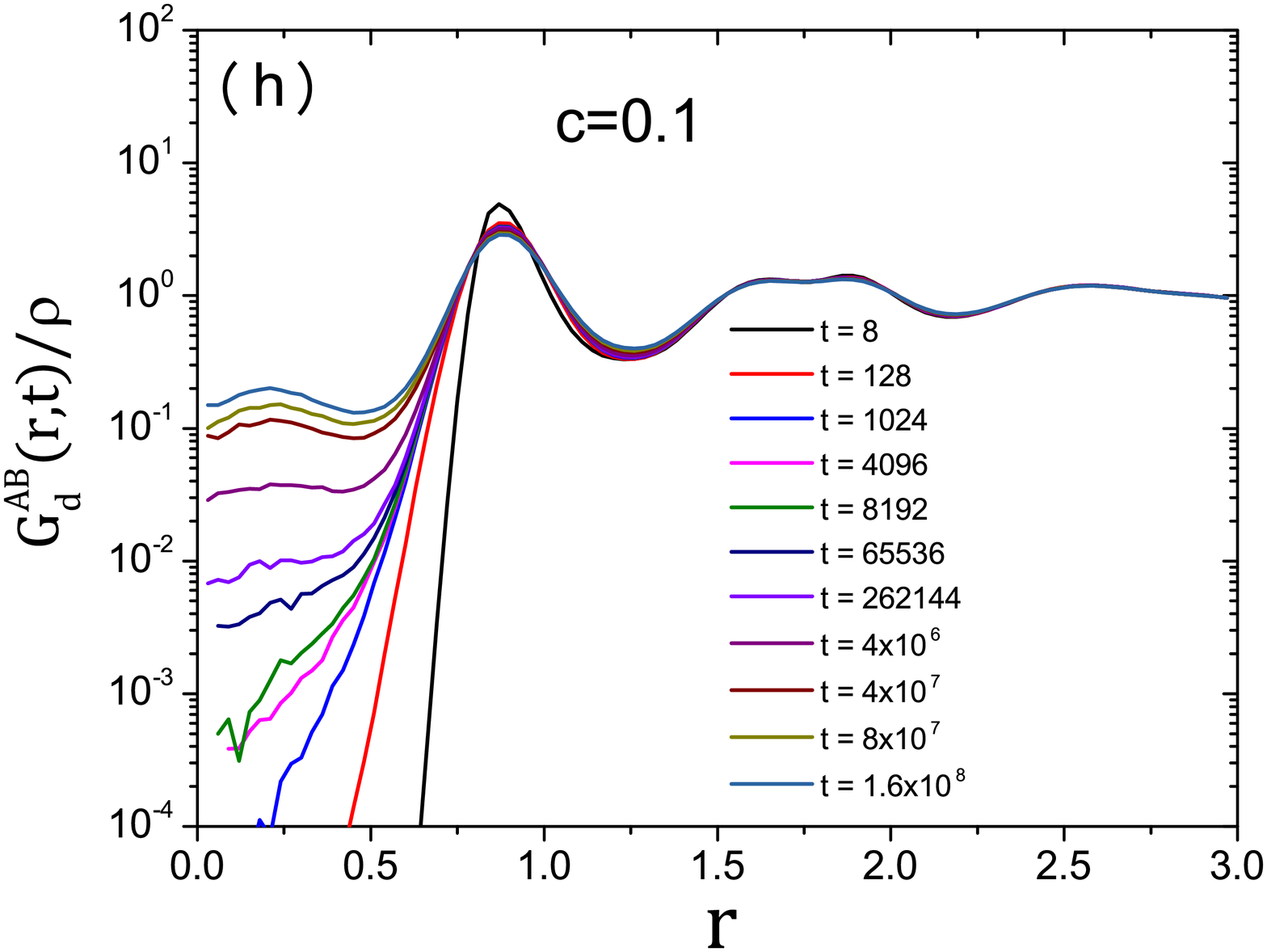}
\includegraphics[width=0.66\columnwidth]{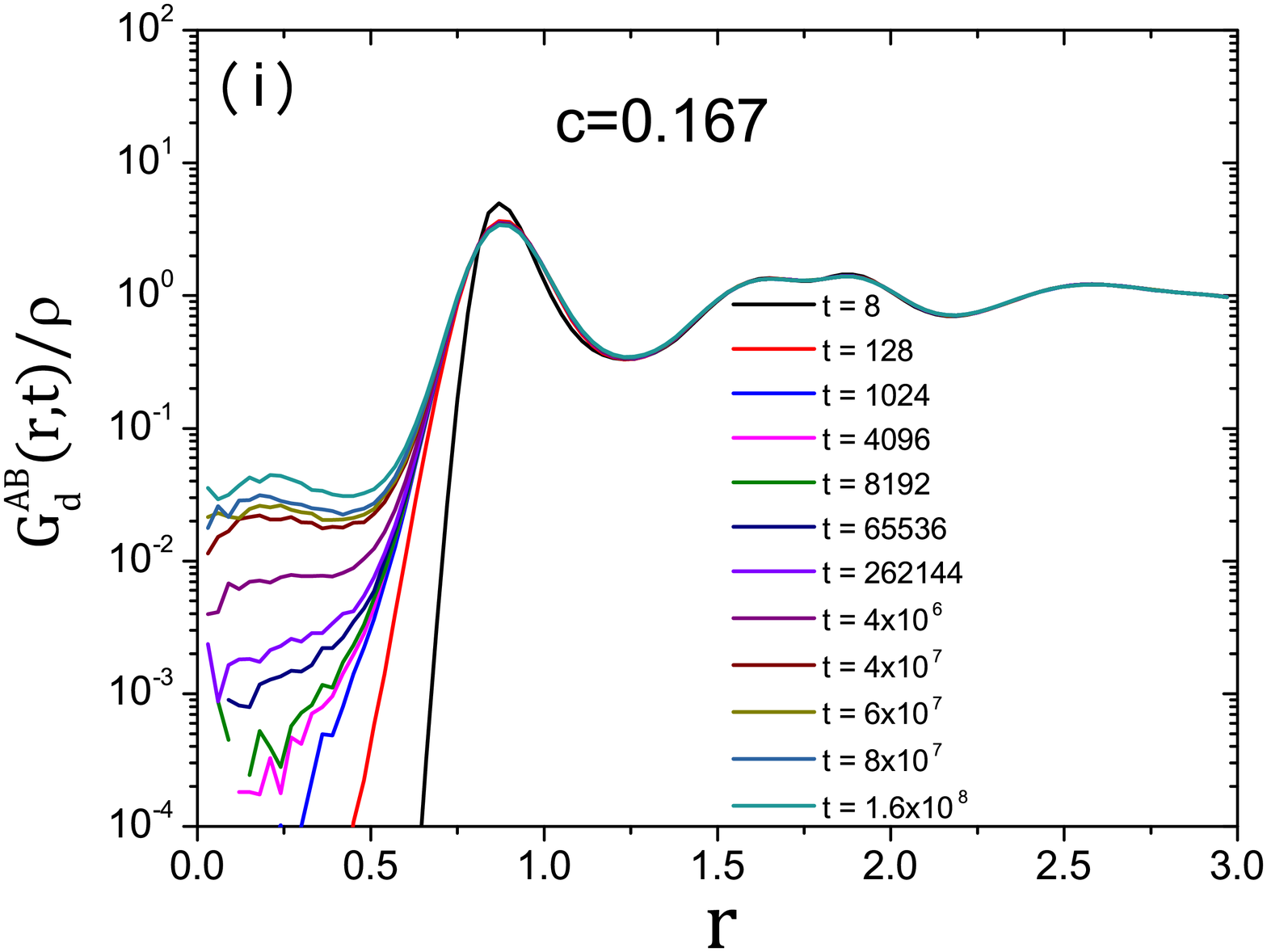}
\caption{
Distinct part of the van Hove correlation function at $T=0.45$ and different values of $c$ for $N=1200$.
(a)-(c): A-A correlation, (d)-(f): B-B correlation, (g)-(i): A-B correlation.
}
\label{fig:vanHove_distinct}
\end{figure*}

The $r$-dependence of the distinct part of the van Hove function
is shown in Fig.~\ref{fig:vanHove_distinct}.  By definition, $G_{\rm
d}^{\alpha\beta}(r, t=0)$ is the radial distribution function, which is
zero at $r  \lesssim 1 $ and with several distinct peaks corresponding to
the nearest or next nearest distances.  All panels show that the hole
of $G_{\rm d}^{\alpha\beta}(r, t=0)$ at $r \lesssim 1$ is filled
up by other particles at longer times.  For the A-A component, panels
(a)--(c), one observes that, for a fixed $t$, the peak developing at
$r=0$ becomes sharper and the height of the peak becomes lower as the $c$
is increased.  We find a very similar behavior for the B-B component,
panels (d)--(f).  The peaks are slightly higher than those of the A--A
component because the B particles can undergo a hopping motion more easily
since they are smaller than the A particle.  On the contrary, for the A-B
component, panels (g)--(i),  the growth of the peak is somewhat weaker.
These results imply that, in the hopping motion, the particles tend to
be replaced by the same kind of particles.

Note that, although the exchange of two particles of the same type
allows that the particles undergo a diffusive motion at large times,
the underlying structure does not change (i.e., it is like the
diffusion of defects in a crystal). Hence this type of motion will
not give rise to a change in the overlap function. This is different for the exchange of an
A particle with a B particle since an exchange of unlike particles will
lead to a non-equivalent configurations. As mentioned above, we recognize
that in this case the peak at the origin is now much smaller which is
explained by the fact that the size of the A particle is significantly
larger than the one of the B particles and hence it is difficult to
swap the position of the unlike particles. Nevertheless, at sufficiently
long times a small peak can be observed even in this case and hence the
system is indeed able to access new configurations, in agreement with
the findings on the entropy and PEL analysis presented in the main text.

\section{Thermodynamics}

\subsection{Anharmonic contribution for the vibrational entropy}

In this section we show how to evaluate the anharmonic contribution
for $s_{\rm vib}$.  Details of this method are described in
Refs.~\cite{mossa2002dynamics,sciortino2005potential} for the bulk system.
The extension of the method to the pinned system is straightforward.

The anharmonic contribution of the total potential energy $U_{\rm anh}(c, T)$ is given by

\begin{equation}
U_{\rm anh}(c, T) = U(T) - U_{\rm IS}(c, T) - \frac{3}{2}(1-c)N  T,
\label{eq:U_anh}
\end{equation}

\noindent
where $U(T)$ and $U_{\rm IS}(c,T)$ are, respectively, the equilibrium bulk
potential energy and the inherent structure energy of the pinned system.
The last term in Eq.~(\ref{eq:U_anh}) is the energy of the harmonic vibrations of $(1-c)N$
mobile particles.  

Making a $T-$expansion of $U_{\rm anh}(c, T)$ around $T=0$ gives

\begin{equation}
U_{\rm anh}(c, T) = \sum_{k=2} C_k (c) T^k,
\label{eq:U_anh_expand}
\end{equation}

\noindent
where $C_k (c)$ are $T-$independent coefficients. Note that the sum
starts at $k=2$ since the system is completely harmonic in the low
temperature limit.

The anharmonic contribution for the vibrational entropy $S_{\rm anh}(c,
T)$ is given by

\begin{equation}
S_{\rm anh}(c, T) = \int_0^T dT' \frac{1}{T'} \frac{\partial U_{\rm anh}(c, T')}{\partial T'}.
\label{eq:S_anh}
\end{equation}

\noindent
Note that we set $S_{\rm anh}(c, T=0)=0$.

Substituting Eq.~(\ref{eq:U_anh_expand}) to Eq.~(\ref{eq:S_anh}), we obtain
\begin{equation}
S_{\rm anh}(c, T) = \sum_{k=2} \frac{k}{k-1} C_k(c) T^{k-1}.
\label{eq:S_anh2}
\end{equation}

\begin{figure}
\includegraphics[width=0.95\columnwidth]{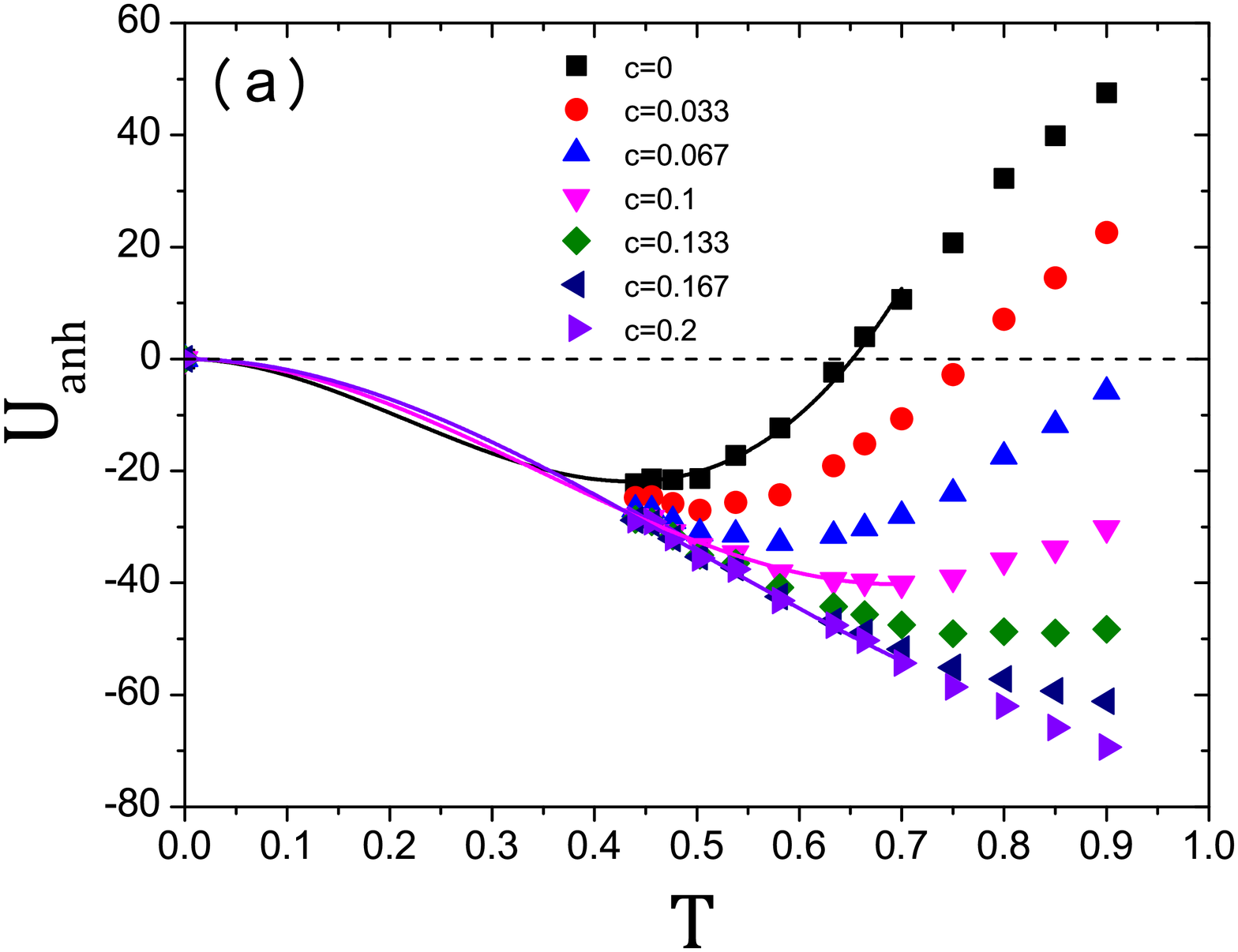}
\includegraphics[width=0.95\columnwidth]{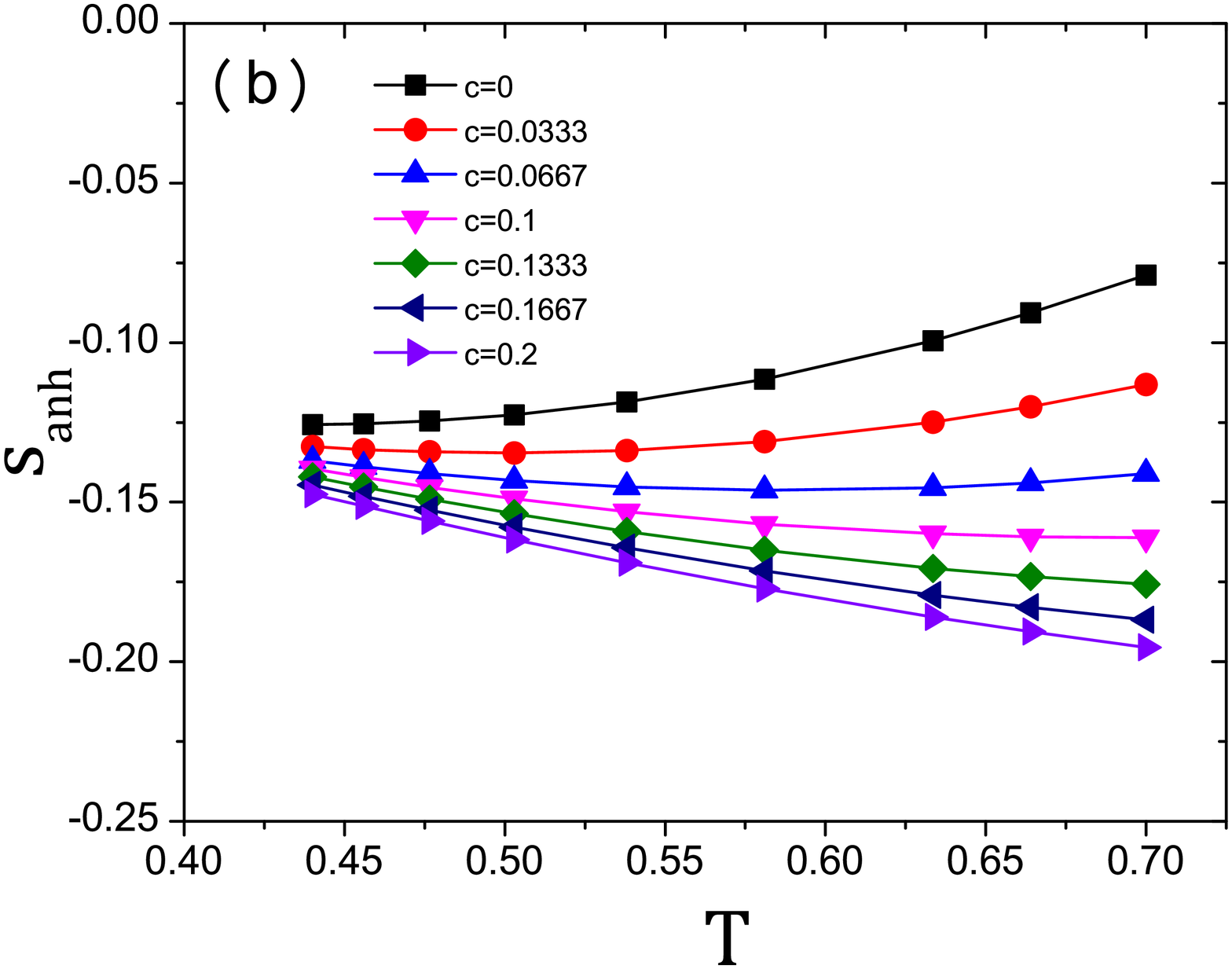}
\caption{\label{fig:anharmonic} 
(a) Temperature dependence of the anharmonic potential energy
$U_{\rm anh}(c,T)$ for several values of $c$'s.  (b) Temperature dependence
of the  anharmonic contribution of the vibrational entropy $s_{\rm
anh}(c,T)=S_{\rm anh}(c,T)/(1-c)N$ per a particle for several values of $c$.
The data in (a) (b) are from $N=1200$.
}
\label{fig:anh}
\end{figure}

\noindent
To evaluate $S_{\rm anh}$ we have used the following numerical procedure:
First we use the simulations to obtain $U_{\rm anh}(c, T)$ and then the
coefficients $C_k (c)$'s are obtained by fitting $U_{\rm anh}(c, T)$
to a polynomial function of $T$.  In practice we use the first two terms, $C_2 (c)$
and $C_3 (c)$, for this fitting~\cite{mossa2002dynamics}. Finally $S_{\rm anh}(c, T)$
is evaluated using Eq.~(\ref{eq:S_anh2}). It is clear
that the signs of $C_k (c)$'s determine the sign of $S_{\rm anh}(c,T)$.
In Fig.~\ref{fig:anh}(a), the simulation data of $U_{\rm anh}(c,T)$
are plotted for several values of $c$.  The solid lines are fits 
down to $T=0$. The resulting $s_{\rm anh}(c,T)=S_{\rm anh}(c,T)/(1-c)N$
are shown in Fig.~\ref{fig:anh}(b).  We find that $s_{\rm anh}(c,T)$
is small and negative for all $c$'s and $T$'s considered, in agreement
with the results from Refs.~\cite{mossa2002dynamics,sciortino2005potential}.

\subsection{Overlap and Finite Size Effects}

\begin{figure*}[htb]
\includegraphics[width=0.66\columnwidth]{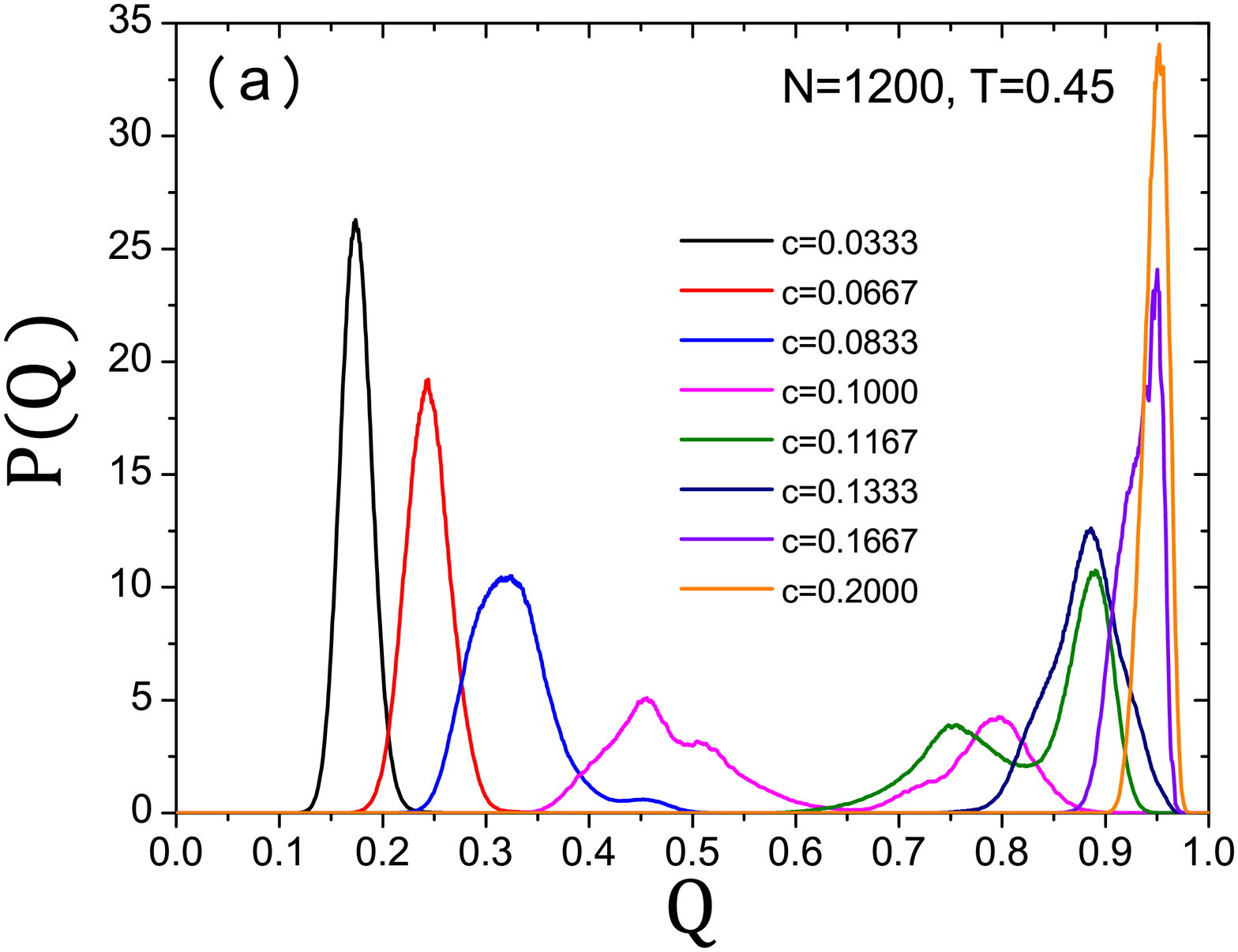}
\includegraphics[width=0.66\columnwidth]{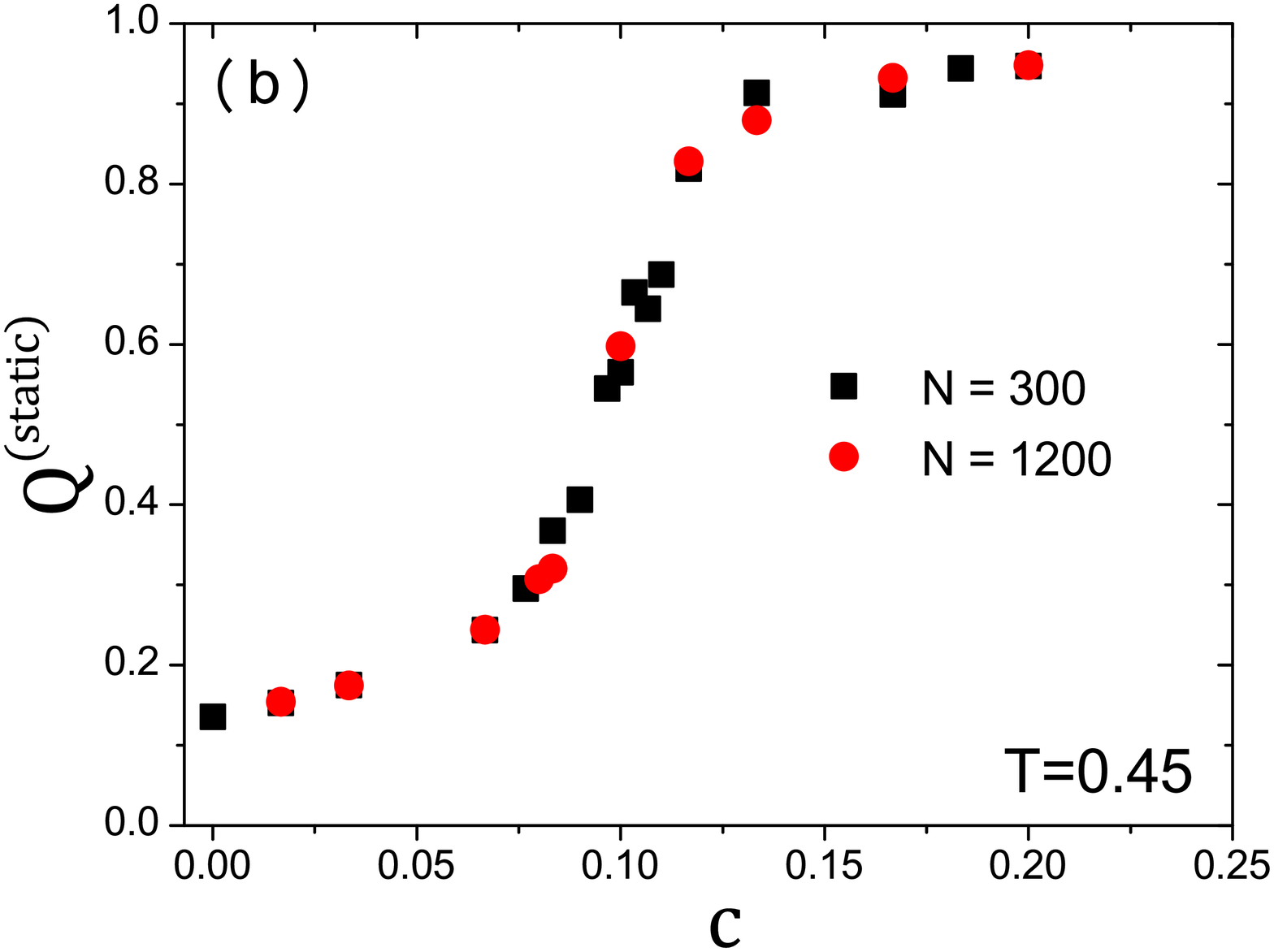}
\includegraphics[width=0.66\columnwidth]{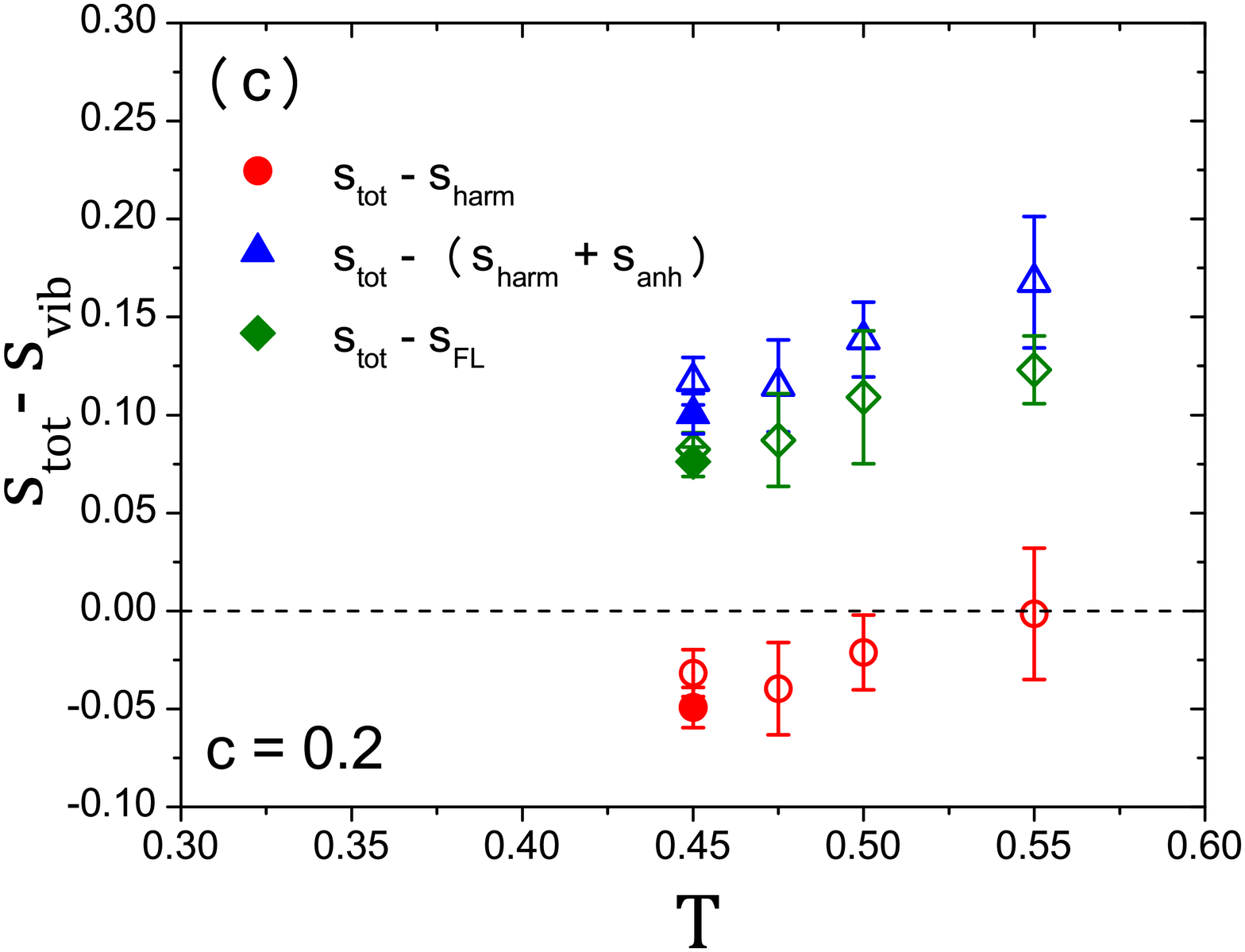}
\caption{
(a) Distribution function $P(Q)$ of the overlap for different
concentration of pinned particles.  $T=0.45$ and $N=1200$. (b)
$c-$dependence of the average overlap for $N=300$ and $N=1200$ at
$T=0.45$.  (c) $T-$dependence of the configurational entropy along
$c=0.2$. The open and filled symbols are for $N=300$ and $N=1200$,
respectively.
}
\label{fig:overlap}
\end{figure*}

In Fig.~\ref{fig:overlap}(a) we show the distribution $P(Q)$ of
the overlap $Q_{\alpha\beta}$ between two configurations $\alpha$ and
$\beta$ which is defined by 
\begin{equation}
Q_{\alpha\beta}=\frac{1}{N}\sum_{i,j}^{N}\theta(a-|\mathbf{r}_i^\alpha-\mathbf{r}_j^\beta|), 
\end{equation}

\noindent
where $\theta(x)$ is the Heaviside function, $\mathbf{r}_i^\alpha$
is the position of the $i$-th particle in the configuration $\alpha$,
and the length-scale $a$ is set to $0.3$~\cite{kob2013probing}. For
small $c$ we see a single peak at small $Q$ which shows that for weak
pinning the overlap between two typical equilibrium configurations is
small. For intermediate $c$ we find a pronounced double peak structure
indicating the coexistence of states that have a low overlap and states
that have a high overlap with each other. If $c$ is increased further
one finds only a single peak at large $Q$, i.e., most configurations
are very similar.  These results are for $N=1200$ particles and
are qualitatively very similar to the one for $N=300$ presented in
Ref.~\cite{ozawa2015equilibrium}. To show the $N-$independence more
quantitatively, we present in Fig.~\ref{fig:overlap}(b) the $c-$dependence
of the mean overlap $Q^{\rm (static)}$ which is given by the first moment
of $P(Q)$. We see that within the accuracy of the data the curves for
$N=300$ coincide with the one for $N=1200$.  We also observe that the
$N$-dependence of $s_{\rm tot}-s_{\rm vib}$ at the lowest temperature
is small (see Fig.~\ref{fig:overlap}(c)).  This result is somewhat
surprising since one expects noticeable finite size effect in the vicinity
of the ideal glass transition due to some growing static correlation
lengths~\cite{fullerton2014investigating,takahashi2015evidence} and,
thus of the $N$-dependence of the order-parameter $Q^{\rm (static)}$.  One
possibility to rationalize the absence of such finite size effects is that
the finite size exponents is very small in the temperature range which
we explore~\cite{cammarota2013random}.  In order to verify this, one must
study lower temperatures with more efficient algorithms or consider other
model systems~\cite{ninarello2017models,berthier2017configurational}.

On this point, we note that our results are not very sensitive to the
temperature, see e.g.~Fig.~\ref{fig:overlap}(c) for the difference
$s_{\rm tot}-s_{\rm vib}$, and therefore we expect them to be relevant
also at lower temperatures.

\section{Potential energy landscape}

\subsection{Snapshot}

\begin{figure*}[htb]
\includegraphics[width=0.75\columnwidth]{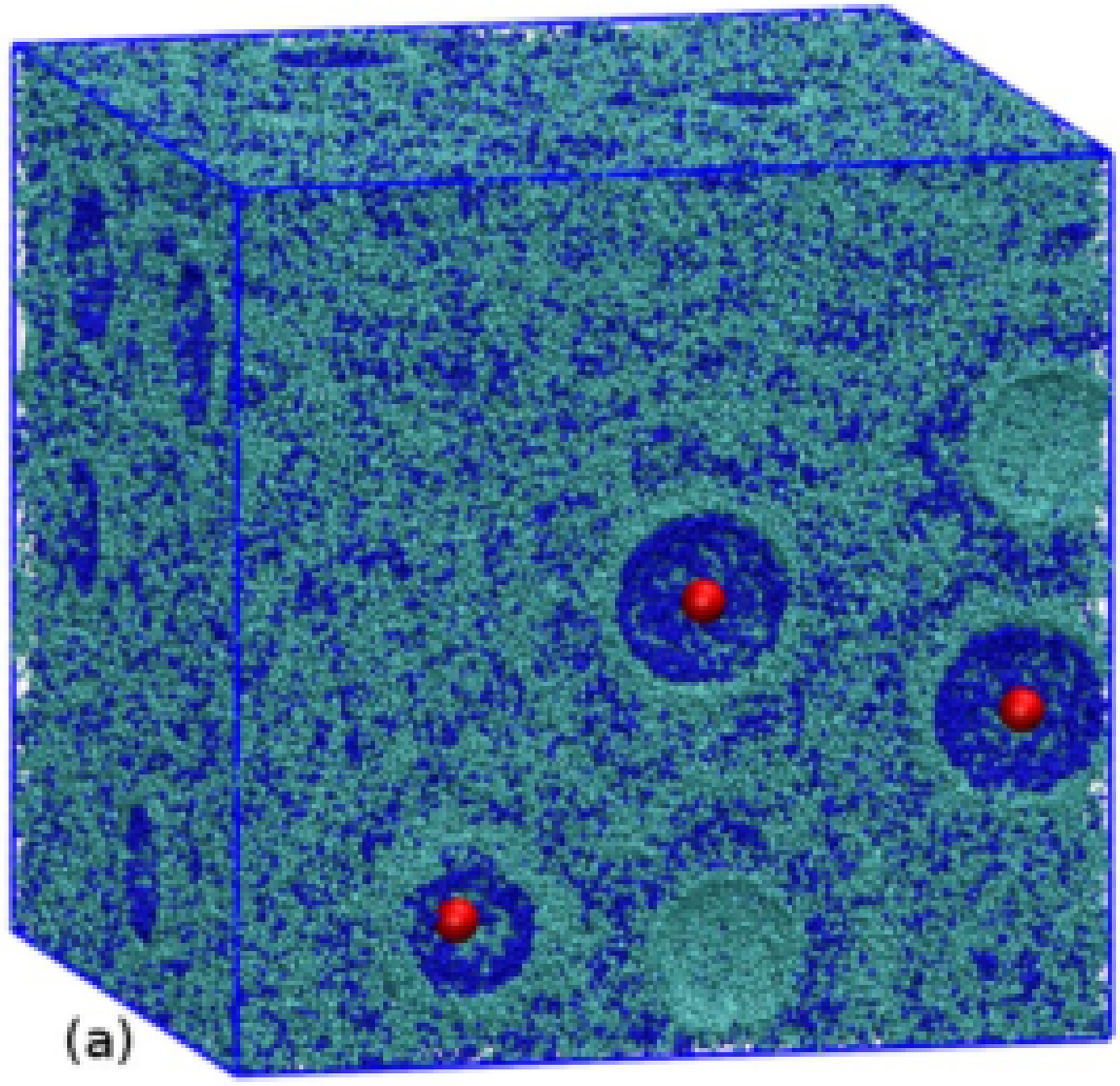}
\includegraphics[width=0.64\columnwidth]{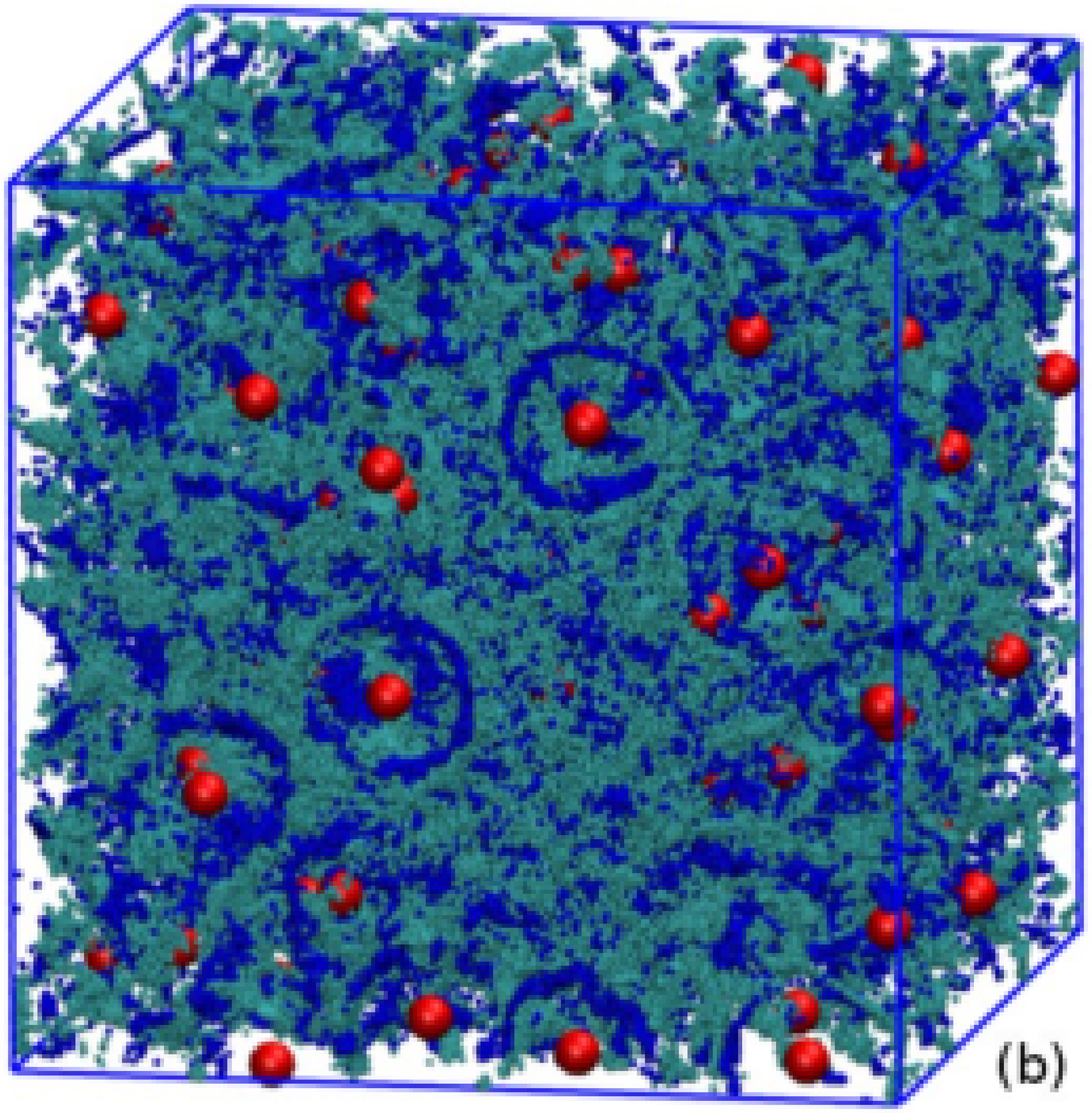}
\includegraphics[width=0.65\columnwidth]{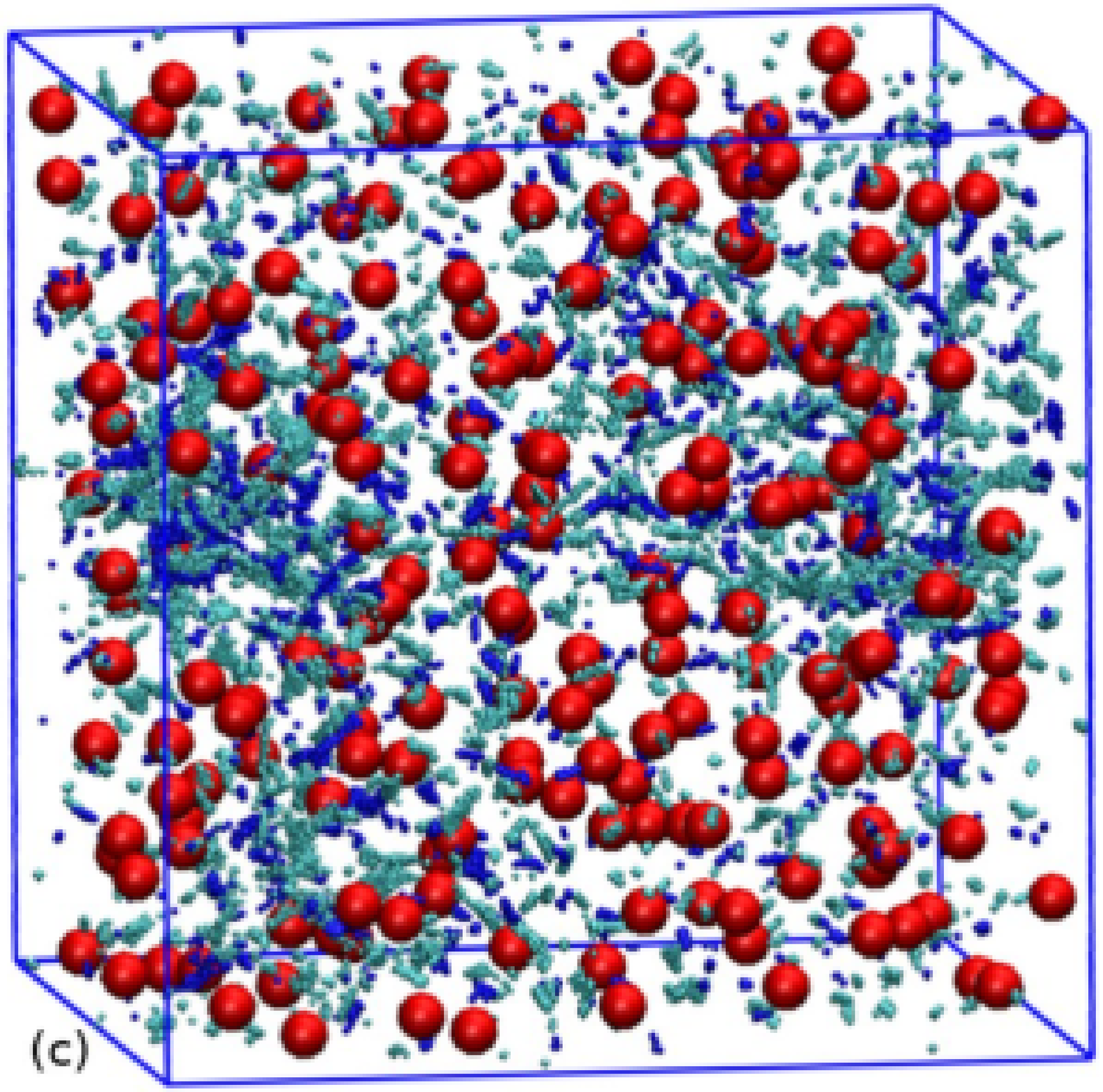}
\caption{Snapshot of IS for $N=1200$, $T=0.45$. The size of the box is
10.0, pinned particles are shown in red, and mobile A and B particles
are shown in blue and gray, respectively.  (a) $c=0.0167$ (fluid), (b)
$c=0.08$ (fluid close to the ideal glass transition), and (c) $c=0.1667$
(glass).
}
\label{fig:snapshot}
\end{figure*}

In Fig.~4 of the main text we show the superposition of snapshots of
IS configurations for $N=300$ particles. In Fig.~\ref{fig:snapshot}
we present similar data for $N=1200$. Panel~(a) shows the snapshots for
$c=0.0167$, i.e., the case where the system is still in the liquid-like
regime. We see that the position of the particles basically fill out
all the accessible space, with the exception of the region around the
pinned particles. For $c=0.08$, i.e., slightly below the transition,
some empty regions can be spotted, indicating that now the particles
can no longer access the whole space, panel (b). If the concentration
is increased to $c=0.167$, i.e., a state that is deep in the glass state,
most of the IS are very similar and hence the available configuration
space has become very small, panel (c). However, even in this case we
find that there are small clusters of particle positions that indicate
that a glass state is composed of many different IS, which is thus the reason
for having a non-trivial entropy.

\end{document}